# An HCAI Methodological Framework (HCAI-MF): Putting It Into Action to Enable Human-Centered AI

Wei Xu, *Senior Member, IEEE,* Zaifeng Gao, Marvin Dainoff

***Abstract*—Human-centered artificial intelligence (HCAI) is a design philosophy that prioritizes humans in the design, development, deployment, and use of AI systems, aiming to maximize AI's benefits while mitigating its negative impacts. Despite its growing prominence in literature, the lack of methodological guidance for its implementation poses challenges to HCAI practice. To address this gap, this paper proposes a comprehensive HCAI methodological framework (HCAI-MF) comprising five key components: HCAI requirement hierarchy, approach and method taxonomy, process, interdisciplinary collaboration approach, and multi-level design paradigms. A case study demonstrates HCAI-MF's practical implications, while the paper also analyzes implementation challenges. Actionable recommendations and a "three-layer" HCAI implementation strategy are provided to address these challenges and guide future evolution of HCAI-MF. HCAI-MF is presented as a systematic and executable methodology capable of overcoming current gaps, enabling effective design, development, deployment, and use of AI systems, and advancing HCAI practice.**

***Keywords*—Artificial intelligence, human-centered artificial intelligence, methodology, human-centered design**

## I. Introduction

AI systems have historically been designed, developed, and deployed with a technology-centered focus. However, concerns about AI's misuse and negative societal impacts have led researchers to advocate for a human-centered AI (HCAI) approach [1], [2], [3], [4]. HCAI prioritizes humans in AI system design and aims to maximize benefits while minimizing harm. It complements technical practices by ensuring AI serves and supports humans rather than replacing or harming them.

Despite its strong conceptual foundation, implementing HCAI remains challenging. A key obstacle is the lack of comprehensive methodologies to bridge the gap between design philosophy and practical execution [5], [6], [7], [8]. This challenge mirrors the early development of human-centered design (HCD) in the 1980s, which initially lacked practical methodologies but evolved through practice to address user experience issues during the computer era [9], [10].

As we enter the AI era, HCAI needs practical frameworks to go beyond its role as a design philosophy and address real-world challenges. Current HCAI frameworks lack effective guidance for practical implementation. To address this, this paper addresses the critical research question: *How can a comprehensive HCAI methodological framework be built to support the adoption of HCAI in practice?*

This paper proposes an HCAI methodological framework (HCAI-MF) based on a literature review and prior research. It includes five key components: an HCAI requirement hierarchy, an approach and method taxonomy, a process, an interdisciplinary collaboration approach, and multi-level design paradigms. A case study highlights the framework's practical value, and the paper discusses its implications, challenges, and actionable recommendations for implementation, along with future directions for its improvement.

This paper makes three key contributions:

- *Proposal of HCAI-MF*: We propose an innovative HCAI methodological framework with five key components.
- *Systematic guidance*: HCAI-MF addresses current weaknesses in implementing HCAI and offers structured, executable guidance for effective practice.
- *Implementation strategy and recommendations*: This paper highlights challenges in current HCAI practices and provides actionable recommendations, including a "three-layer" strategy for adopting HCAI-MF.

By bridging the gap between HCAI design philosophy and methodology, this paper supports the effective design, development, deployment, and use of HCAI systems.

## II. Methods

This study used a qualitative approach to review key works in HCAI methodology. A *literature review* across databases like Google Scholar, ResearchGate, ACM Digital Library, IEEE Xplore, and Springer Link established a comprehensive foundation for the study, using keywords such as human-centered AI (HCAI), HCAI methodology, HCAI guiding principles, HCAI framework, and human-AI interaction.

We included peer-reviewed, English-language publications related to HCAI methods, frameworks, principles, design, method, process, and practice since 2018. The search yielded over 200 works, of which 125 were retained after abstract screening. Authors screened independently and resolved any disagreements through discussions.

A *gap analysis* was performed by mapping reviewed works, identifying methodological shortcomings or underexplored areas. This informed a *framework development process*, which followed an iterative, expert-informed approach: drafting the HCAI Methodological Framework (HCAI-MF) with five core components.

To demonstrate applicability, the HCAI-MF was applied in a case study on human–autonomous vehicle co-driving. Following application, a final *gap analysis* was conducted

*(Corresponding authors: Wei Xu; Zaifeng Gao)*

Wei Xu is with the Department of Psychology and Behavioral Sciences and the Center for Psychological Sciences at Zhejiang University, China (email: weixu6@yahoo.com). Zaifeng Gao is with the Department of Psychology and Behavioral Sciences at Zhejiang University, China (email; zaifengg@zju.edu.cn. Marvin Dainoff is with the Department of Psychology, Miami University, USA (email: dainofmj@miamioh.edu).

as the initial review to evaluate potential challenges for implementing the HCAI-MF, producing actionable recommendations and directions for future improvement.

## III. Needs for a human-centered AI methodological framework

### *A. Human-Centered AI (HCAI)*

AI technology is created to benefit humanity, but it can also cause harm if poorly developed or misused. Research identifies several AI limitations, including vulnerability, bias, lack of explainability, absence of causal models, autonomy-related issues, and ethical concerns [11], [12]. Based on rigorous processes, including data verification, standardization, and continuous updates [13], [14], databases like the AI Accident Database, which tracks incidents such as autonomous vehicle accidents and facial recognition errors, have recorded over a thousand cases [15]. Similarly, the AIAAIC database reports a 26-fold increase in AI abuse incidents since 2012 [16].

Many AI projects fail [17], often due to insufficient focus on "human factors" such as poor user experience, lack of explainability and controllability, unmet user needs, and neglect of ethical considerations. These challenges underline the importance of adopting a human-centered approach to the design, development, and deployment of AI systems.

A human-centered approach addresses the limitations of traditional AI methods. Researchers like Shneiderman [2] and Xu [1] have introduced systematic human-centered AI (HCAI) frameworks, focusing on human values, needs, knowledge, and roles throughout AI design, development, deployment, and use. The goal of HCAI is to create AI systems that augment human capabilities and improve well-being, rather than replace them, designing safe, trustworthy, and reliable AI systems [1], [2]. This approach has been further developed in areas like human-centered computing [18], human-centered explainable AI [19], trustworthy AI [20], human-centered machine learning [21], and humanistic AI design [22].

In 2019, Xu [1] proposed one of the earliest HCAI frameworks, the "human factors-technology-ethics triangle", which integrates three perspectives: technology, human factors, and ethics (Figure 1). From the technology perspective, HCAI combines human and machine intelligence to enhance human capabilities with advanced AI technology. The human factors perspective emphasizes designing AI systems controllable, useful, usable, and explainable. The ethics perspective addresses issues like human values, fairness, and privacy. HCAI emphasizes a systematic approach that fosters synergy and complementarity among these perspectives. An HCAI solution is to achieve balance across all three aspects. To achieve its goals, HCAI practice needs interdisciplinary collaboration, ensuring a holistic approach to HCAI practice.

### *B. Search for an HCAI Methodological Framework*

While the conceptual foundations of HCAI are well-explored, practical implementation remains underdeveloped [5], [23], [24], [25]. A major challenge is the lack of comprehensive methodologies to guide its adoption [6], [7], [8]. Researchers have called for robust HCAI methodologies and proposed frameworks addressing areas, such as HCAI design and evaluation solutions [4], interdisciplinary collaboration [8], requirement gathering [26], organizational maturity [27], design processes [28], and service science [29].

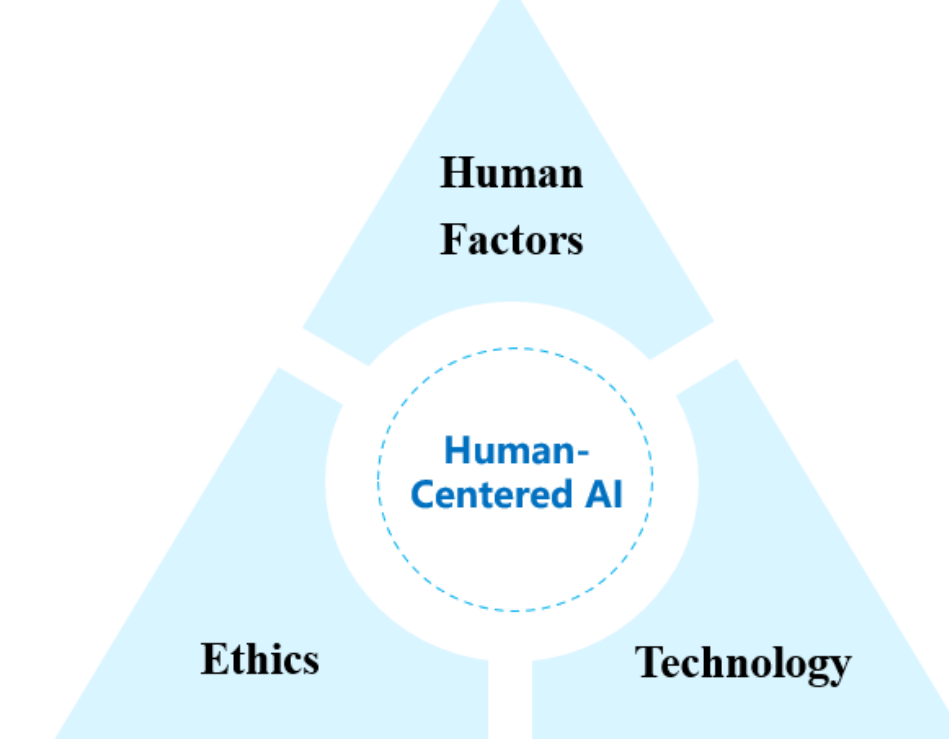

**Fig. 1** The "human factors-technology-ethics triangle" HCAI framework [1]

Many HCAI frameworks focus on specific aspects but often lack comprehensive methodological guidance for practice. For instance, the human control-automation framework emphasizes balancing human control with automation [2], while the human factors-technology-ethics triangle framework integrates ethical design, technology, and human factors [1]. Other frameworks, like ethically aligned design [30] and responsible AI design [31], prioritize ethical principles, while the human-centered requirements framework focuses on managing human-centered requirements in AI design [32].

A methodological framework typically includes design philosophy, guiding principles, requirements, approaches, methods, processes, and design paradigms. However, current HCAI frameworks lack key methodological elements, hindering effective HCAI operationalization. To overcome these gaps and advance HCAI adoption, this paper introduces a comprehensive HCAI methodological framework (HCAI-MF) to provide structured, practical guidance for applying HCAI in practice. The contributions are outlined in Tabel 1.

## IV. An HCAI Methodological Framework

As shown in Figure 2, HCAI-MF includes five components:

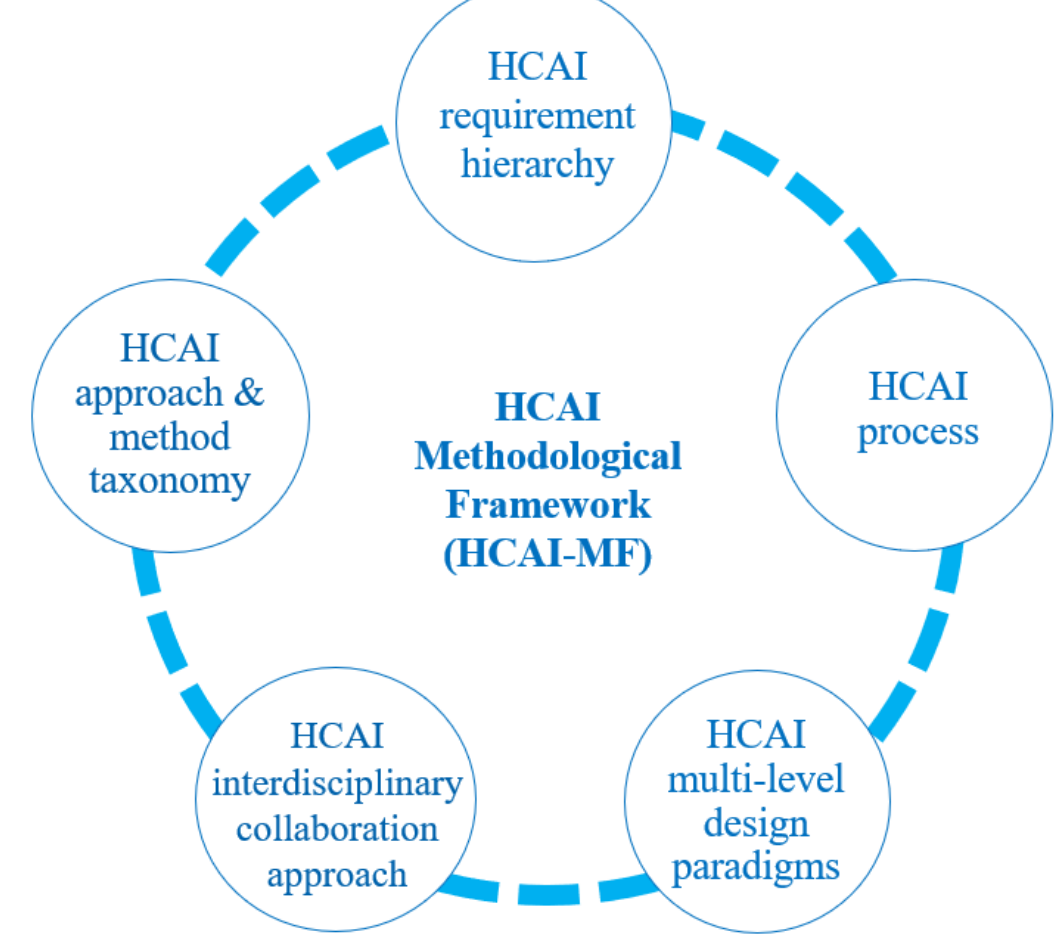

**Fig. 2** An HCAI methodological framework

Table 1 The methodological gaps in current HCAI practice and the contributions of this study

| | Definition | Challenges in current HCAI practice | Contributions of this study |
|---|---|---|---|
| Overall HCAI methodological framework | An HCAI methodological framework is a structured, systematic approach that defines guiding principles, requirements, approach/methods, processes, and design paradigms used to conduct research, analyze, design, and develop AI systems, achieving desired HCAI objectives | • HCAI are well-explored with many HCAI frameworks/concepts proposed, but practical implementation remains underdeveloped [5], [23], [24]<br>• There is a lack of comprehensive methodologies to bridge the gap between HCAI concepts and practice [5], [6], [7], [8] | • Propose a comprehensive HCAI-MF with five key components included<br>• Address weaknesses in HCAI practice and offer structured, executable guidance<br>• Identify challenges in HCAI practices with actionable recommendations, including a "three-layer" strategy for adopting HCAI-MF |
| HCAI guiding principle, design guideline, and requirement | • HCAI guiding principles are high-level, strategic sets of rules industry-wide<br>• HCAI design guidelines are tactical, domain-based high-level requirements<br>• HCAI requirements are measurable requirements, ensuring systems meet HCAI principles and design guidelines | • Many existing HCAI guiding principles often overlap conceptually<br>• Many HCAI guidelines are too broad (e.g., ethics), being too abstract for practices [33]<br>• Unclear relations between HCAI principles, guidelines, and requirements, making it hard to define HCAI requirements [26], [34] | • Consolidate existing HCAI guiding principles into eight distinct HCAI guiding principles<br>• Bridge the gap between strategic goals and practical implementation, making HCAI guiding principles and design guidelines more actionable<br>• Propose an HCAI requirement hierarchy to guide projects and derive product HCAI requirements |
| HCAI approach & method | HCAI approach & method are implementation strategies and procedures that guide project work toward HCAI design goals | Although HCAI design and evaluation solutions have been proposed [4], there is still no comprehensive framework to drive strategy and scalability for operationalize HCAI. Overall, HCAI methods appear to lag in helping HCAI practice HCAI [5], [35], [36] | • Propose a structured and scalable HCAI approach & method taxonomy, integrating key HCAI strategies and activities across AI lifecycle<br>• Align key human factors (e.g., value, need, experience) and HCAI guiding principles across HCAI approaches and methods |
| HCAI process | HCAI process is a structured series of activities, steps, or stages designed to guide HCAI practice across the entire lifecycle of AI system development | • There is no comprehensive HCAI process; HCAI activities are limited [28], [37]<br>• Design relies on traditional software practice, HCAI practice starts too late [35]<br>• HCAI gets lost [6], [8]; HCAI principles & guidelines are not consistently applied [35] | • Propose an HCAI process to integrate human-centered approach across entire AI lifecycle<br>• Offer high-level guidance on integrating HCAI methods and activities throughout the lifecycle.<br>• Addresses fairness, accountability, privacy, and bias at every stage of the lifecycle. |
| HCAI interdisciplinary collaboration | HCAI practice requires interdisciplinary collaboration across multiple disciplines in designing, developing, deploying, and using AI systems towards HCAI goals | Multi-disciplinary participation is recognized [4], but in the reality, miscommunication and misaligned goals in HCAI practice exist; human-centered professionals are either involved too late or not involved [8], [38], [39] | • Propose interdisciplinary collaboration strategy and process, including goal setting, team building, co-design approach, HCAI process adoption, communication and decision-making flows, and success measurement |
| HCAI design paradigm | HCAI design paradigm is a specific lens to determine the scope, perspective, and approaches for HCAI practice | Current HCAI practice is less comprehensive and systematic with a focus on individual human-AI systems, lacking diverse paradigms for comprehensive HCAI solutions [11], [40] | • Propose multi-level HCAI design paradigms to holistic ecosystem & sociotechnical perspectives<br>• The multi-level HCAI design paradigms help define HCAI practical strategies in early stages |

HCAI requirement hierarchy, approach and method taxonomy, process, interdisciplinary collaboration approach, and multi-level design paradigms. This section details each of the five components and their contributions to achieving HCAI goals.

### *A. HCAI Requirement Hierarchy*

Current HCAI practice faces several challenges, including ineffective methods for gathering HCAI requirements [26]. Although many HCAI guiding principles, such as explainable AI [19], ethical AI [41], fair AI [42], trustworthy AI [20], and responsible AI [43], have been proposed, they often overlap conceptually. For instance, ethical AI, fair AI, and responsible AI share significant similarities. Additionally, existing HCAI design guidelines often focus on broad concepts like human values, ethics, and privacy, which are too abstract for practice [33]. Another issue is the unclear relationship between HCAI guiding principles, design guidelines, and product requirements, making it challenging for projects to establish HCAI requirements as a foundation for HCAI systems.

To tackle these challenges, we consolidated existing HCAI guiding principles into eight distinct HCAI guiding principles based on a review of key studies (Table 2) [34], [44], [45], [46], [47], [48]. These guiding principles form the foundation of the HCAI design philosophy and aim to achieve HCAI design goals, though they may evolve over time.

We also propose an HCAI requirement hierarchy (HCAI-RH), a structured framework that organizes guiding principles, design guidelines, and product requirements into a hierarchical relationship (Figure 3), with each level defined as follows:

- *HCAI design goals:* Broad rules for AI systems, focused on enhancing human capabilities and performance rather than harming or replacing them.
- *HCAI guiding principles*: High-level, strategic set of rules adopted industry-wide in guiding practices to HCAI goals.
- *HCAI design guidelines:* Tactical, domain-specific rules outlining high-level requirements for AI systems, aligned with HCAI principles. These may follow industry standards or be customized for specific organizations.
- *Product detail requirements:* Specific and measurable requirements, including functional and non-functional (e.g., user performance), ensuring the AI system meets HCAI guiding principles and design guidelines.

The HCAI requirement hierarchy, as shown in Figure 3, illustrates a structured framework with three key features:

- *Means-end relationships*: Lower-level elements serve as "means" to achieve the "ends" of higher-level objectives.
- *1-to-many mapping:* Each upper-level element links to multiple lower-level elements for comprehensive coverage.
- *Goal-driven directionality*: A top-down structure guides

Table 2 HCAI guiding principles

| # | Design Principles | Definition and HCAI Goals |
|---|---|---|
| 1 | Transparency and Explainability [19], [49] | Provide explainable and understandable AI output for humans to enhance human trust and empower informed decisions. |
| 2 | Human Control and Empowerment [50], [51] | Allow humans to understand, influence, and control AI behavior when necessary; Ensure AI aligns with human needs. |
| 3 | Ethical Alignment [41], [47], [52], [53], [54] | Develop AI in alignment with ethical and societal norms to preserve human values and privacy, foster trust, and minimize harm. |
| 4 | User Experience [55], [56] | Create interactions that are engaging, intuitive, accessible, and aligned with user expectations. |
| 5 | Human Augmentation & Human-Led Collaborative Interaction [11], [57] | Design AI to enhance human abilities and human-led collaborative interaction to enhance productivity and effectiveness. |
| 6 | Safety & Robustness [58], [59] | Prioritize human safety, maintain reliability in diverse scenarios to ensure resilience and reduce potential risks to humans. |
| 7 | Accountability [60], [61] | Ensure the implementation of responsible AI mechanisms that establish clear accountability for humans (e.g., operators, developers) in relation to AI actions |
| 8 | Sustainability [62], [63] | Develop AI to support environmental, social, and economic well-being to prioritize human well-being while aligning with sustainability. |

projects from strategic HCAI goals and guiding principles to design guidelines and actionable detailed requirements.

This hierarchical framework ensures that product requirements are rooted in HCAI design goals and guiding principles, supported by human-centered requirement-gathering methods (e.g., user interviews). It bridges the gap between strategic product goals and practical implementation, making HCAI guiding principles and design guidelines more actionable and practical. Project teams can leverage this framework to effectively derive HCAI-based requirements.

## *B. HCAI Approach and Method Taxonomy*

Research shows that current HCAI practices lack actionable and effective methods. Non-AI professionals, like human-computer interaction (HCI) and human factors practitioners, often struggle with activities such as generating interaction ideas, designing conceptualization, prototyping, and testing [64], [36]. To address these challenges, researchers stress the need for HCAI design and evaluation solutions, and the need for improving existing methods [4], [65], [66], [67].

To further support HCAI adoption, HCAI-MF introduces an HCAI approach and method taxonomy (HCAI-AMT), providing a comprehensive and scalable framework to operationalize HCAI. The categorization was defined using an informal hybrid card-sorting exercise, combining closed sorting (based on prioritized human factors and HCAI guiding principles; see Columns 4 and 6 in Appendix A) and open sorting (based on literature review and multidisciplinary methods). As a result, the comprehensive taxonomy organizes HCAI approaches and methods into five categories: HCAI strategy, HCAI-based computing, interaction technology and design, human controllability, and AI governance (Figure 4).

To illustrate the values of the taxonomy, Appendix A presents 16 selected HCAI approaches and methods, outlining definitions and key HCAI activities across the AI lifecycle. While some methods enhance existing technology-centered approaches, such as human-centered explainable AI and machine learning, others, adapted from non-AI fields, complement current AI practices. This scalable taxonomy is designed to be expandable, allowing for additional approaches and methods to be integrated as the field evolves.

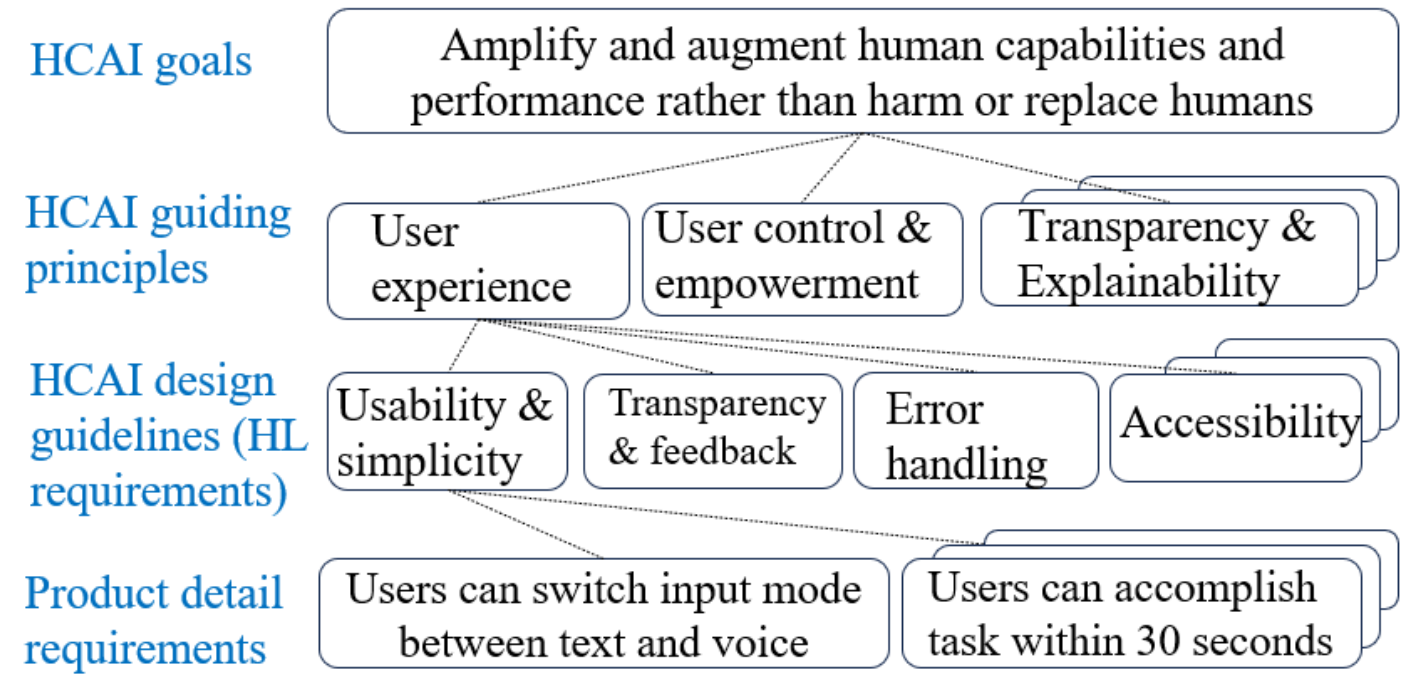

**Fig. 3** Illustration of HCAI requirement hierarchy

Importantly, Figure 4 (the labels within the circle) and Appendix A (the "human factors prioritized" column) highlight how these 16 approaches and methods focus on key human factors, such as human needs, values, abilities, roles, and controllability, throughout the AI lifecycle, in alignment with HCAI guiding principles (see the final column of Appendix A), thereby reinforcing and contributing to the realization of these HCAI guiding principles. Thus, this comprehensive taxonomy embodies HCAI's human-centered focus, guiding practices to prioritize these human factors in achieving HCAI goals

The taxonomy provides three key benefits:

- *Goal-oriented design*: It clearly identifies the human factors to be focused on and aligns them with HCAI guiding principles using tailored approaches and methods.
- *A comprehensive and scalable framework:* The taxonomy encompasses a diverse selection of approaches and methods that can be customized for various HCAI practices and scaled to accommodate future developments.
- *Practical and actionable guidance:* It directly addresses current HCAI challenges through adopting these approaches and methods, enhancing the effectiveness of practices while bridging gaps across disciplines.

Thus, this taxonomy serves as a useful framework to guide current HCAI practices and support continuous refinement.

## *C. HCAI Process*

Effective processes are crucial to guiding projects toward achieving design goals. Research shows that current HCAI practices face challenges from the process perspective. For example, Hartikainen et al. [5] found that early-phase AI design still relies on traditional software practices. HCAI practitioners are often involved after requirements are defined, limiting their influence [35]. Yang et al. [38] highlighted challenges like prototyping and handling unpredictable AI behavior. Bingley et al. [6] noted that while developers consider ethics and security, the broader HCAI perspective often gets less attention. Mazarakis et al. [8] found issues like separating HCAI from technical work and relying on business clients instead of end users. Validation often overlooks AI biases, and HCAI principles are not consistently applied [35].

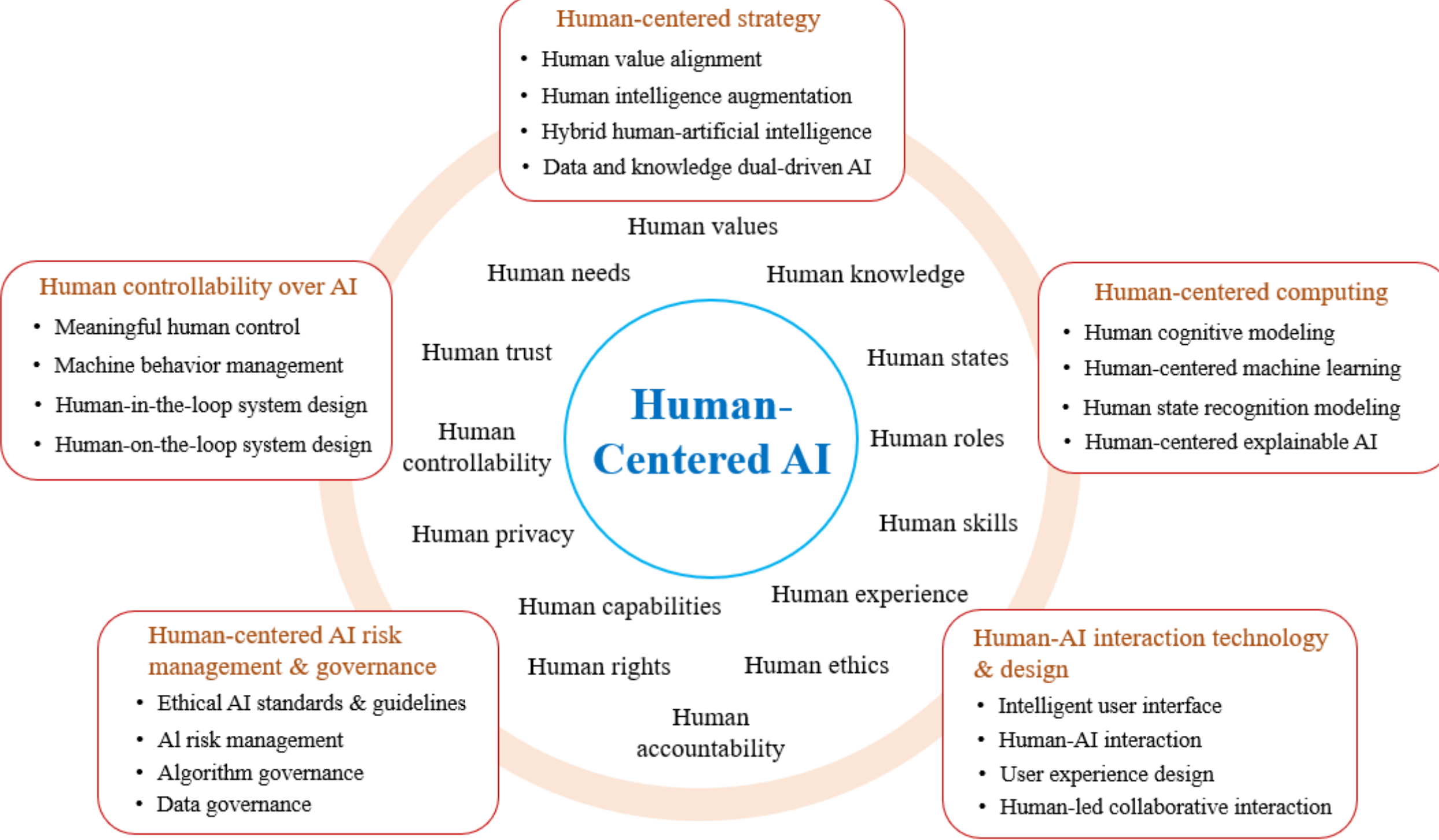

**Fig. 4** HCAI approach and method taxonomy

Researchers have proposed HCAI-based processes to address existing gaps. Some suggest maturity models to improve HCAI practices [23], while others focus on specific stages. For example, Ahmad et al. [26] applied HCAI principles to requirements gathering, Cerejo [28] emphasized collaboration between AI and HCI professionals, and Battistoni et al. [36] adapted a human-centered design (HCD) process. However, a comprehensive HCAI process has yet to be developed. Like the HCD process, an HCAI process should span the entire AI lifecycle, guided by HCAI guiding principles. It must prioritize human needs, values, and roles at every stage, including post-release activities like ethical governance, monitoring AI behavior, and gathering user feedback.

To address the gap in current HCAI practice, we introduce an interdisciplinary HCAI process (Figure 5). It combines the HCD process with a generalized AI lifecycle [93], [94], providing a holistic approach to developing HCAI systems while tackling the unique challenges of AI technology.

- *The HCD Process*: The HCD process is a user-centered, iterative approach focused on addressing human needs and experiences (Figure 5) [93]. It emphasizes understanding problems from the user's perspective, with active user involvement and feedback at every stage. Interdisciplinary collaboration ensures diverse expertise, while prototyping and testing allow for rapid idea exploration and iterative refinement. The process balances user experience with organizational goals, delivering practical outcomes.
- *The AI-Based System Development Lifecycle:* The AI development lifecycle, as illustrated in Figure 5, represents a generalized framework encompassing all stages of the lifecycle of AI-based systems, from problem identification to monitoring and continuous improvement [94].

To illustrate the characteristics of the HCAI process, Figure 5 highlights key activities across the AI lifecycle with six selected HCAI approaches and methods, such as "human value alignment" (Appendix A). These activities ensure that HCAI principles are applied across stages, aligning with human values, addressing ethical concerns, and meeting user needs.

The HCAI process has four key features:

- *Human-centered focus*: Prioritizes human needs, goals, and values, involving users and addressing HCAI-related issues throughout the lifecycle per HCAI guiding principles.
- *Iterative improvement*: Encourages ongoing prototyping, user testing, ethical reviews, AI behavior monitoring, and system refinement based on user feedback.
- *Ethical development*: Addresses ethical issues, such as fairness, accountability, privacy, and bias, across stages.
- *Interdisciplinary collaboration*: Combines interdisciplinary expertise to create human-centered solutions.

The activities shown in Figure 5 only represent selected key HCAI-driven activities for illustration and do not encompass the full range of HCAI activities. The HCAI process offers high-level guidance on integrating HCAI approaches, methods, and activities into the lifecycle. Projects can adapt and optimize the process as needed to suit specific project requirements.

We use large language models (LLMs) as an example to illustrate the value of the HCAI process. LLMs have demonstrated remarkable capabilities. However, from the HCAI perspective, critical issues need to be addressed to ensure LLMs better serve humans [95], [96]. First, lack of

Key HCAI-focused methods and activities across the AI-based system development lifecycle explainability makes it difficult for users to understand or trust their outputs. Second, LLMs often inherit and amplify biases present in training data, leading to ethical issues. Third, the risk of generating harmful or misleading content presents significant ethical and safety challenges. Fourth, aligning LLM behavior with human values remains challenging. Finally, poor human-AI interaction design can hinder effective use, as users may struggle to develop accurate mental models or maintain control. Researchers have explored human-centered solutions to mitigate these issues [97], [98], [99], [100].

| | | | | | | |
|---|---|---|---|---|---|---|
| Human value alignment | Define human value | Ensure fairness in data | Mitigate bias in model selection, training, development | Validate human value | Communicate users clearly | Monitor user feedback |
| User experience design | Define user need/UX goals | User participatory | Define interaction model and user interfaces | Test software/hardware w/ users | Provide user support | Track user feedback |
| Human-centered machine learning | Define user-centric problem | Collect diverse data | Use diverse data, model behavior w/ controlled algorithm | Test for ethics | Gather and integrate user feedback, refine algorithm | |
| Data/knowledge dual-driven AI | Define metrics & scenarios | Extract human knowledge | Define cognitive architectures, train w/ human experts | Validate and integrate data and knowledge | Evolve knowledges/rules | |
| Machine behavior mgmt. | Set behavior goals | Ensure fairness in data | Select explainable model, train model w/ user feedback | Test varied scenarios | Track AI behavior w/ user feedback, refine behavior | |
| AI ethical governance | Establish ethical policies | Mitigate data bias | Choose/train/develop model for fairness/transparency | Perform audits | Communicate users clearly | Monitor ethnical issues |

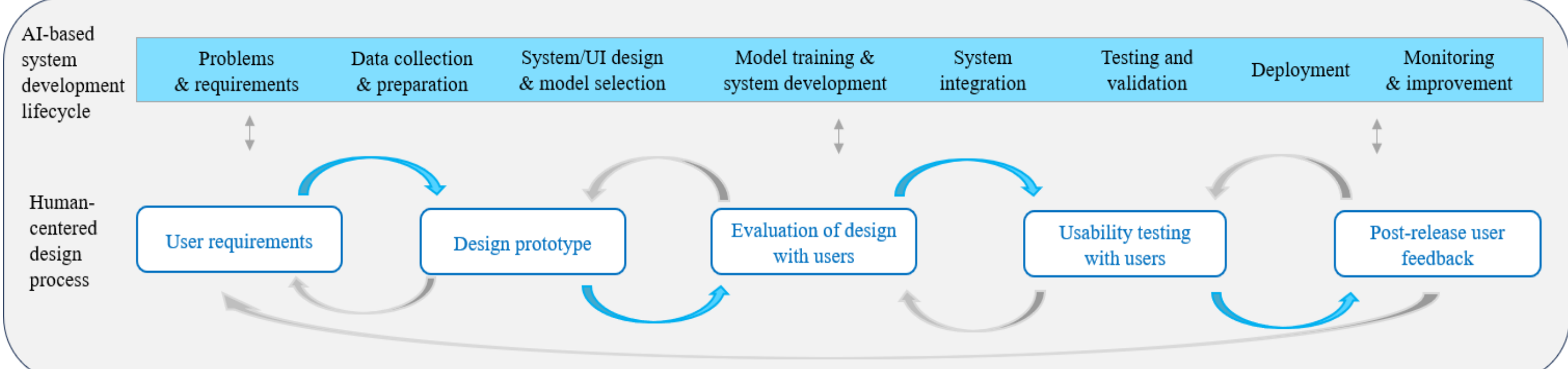

**Fig. 5** HCAI process

One of the underlying causes of these challenges is the insufficient integration of the human-centered approach in the LLM lifecycle [94], [101]. Human-centered LLMs are designed to ensure human-centricity in every stage of the LLM lifecycle. Fede et al. [102] argue that human-centricity is not an optional feature in LLM development but a societal necessity to promote human values. By embedding HCAI principles across stages, LLMs can better serve humans.

We analyzed the root causes by applying the proposed HCAI process to the LLM lifecycle. Appendix B highlights the potential of this HCAI process by comparing a generalized LLM lifecycle with an HCAI-enhanced LLM lifecycle. As shown in Appendix B, the typical activities across the general LLM lifecycle lack a consideration of the human-centered approach. The HCAI-enhanced lifecycle integrates HCAI methods and activities throughout the lifecycle, providing an effective approach to address the HCAI-related challenges.

Developing LLMs is a complex process involving many challenges. However, the HCAI-enhanced process will help projects minimize risks and achieve HCAI design goals.

### *D. HCAI Interdisciplinary Collaboration Approach*

Interdisciplinary collaboration has driven technological advancement. AI has greatly benefited from insights across fields, such as cognitive neuroscience. However, the unique challenges AI poses demand even deeper interdisciplinary approaches to implement HCAI effectively [8], [26], [38].

Our early work shows that AI and computer science fields offer powerful capabilities, while the human factors science fields like HCI and human factors bring valuable insights into HCAI practice [117]. While both fields have strengths and weaknesses, their complementary nature emphasizes the need for collaboration. Interdisciplinary teams have been critical to the success of HCD [118], and HCAI is no exception. Such collaboration is essential to successful HCAI practices. Desolda et al., proposed the disciplines converging in HCAI from the perspectives of design and evaluation solutions [4], such as HCI, AI, software engineering, and ethics. To further address the inefficiency in the collaboration for HCAI, we propose the following approach for organizations:

- *Establish shared goals:* Define a common vision and plan for implementing HCAI.
- *Build interdisciplinary project teams:* Include experts in AI, human-centered design, ethics, and relevant domains.
- *Co-design the AI system:* Involve all stakeholders in the collaborative design process for comprehensive input.
- *Manage conflicts in ethical interpretations:* Establish a shared baseline and ethical boards for universal principles while respecting cross-cultural and disciplinary contexts
- *Adopt the HCAI process & methods:* Apply the HCAI process and methods to enable effective collaboration.
- *Define decision-making frameworks:* Create clear guidelines for collaborative and transparent decisions.
- *Encourage communication:* Promote understanding across

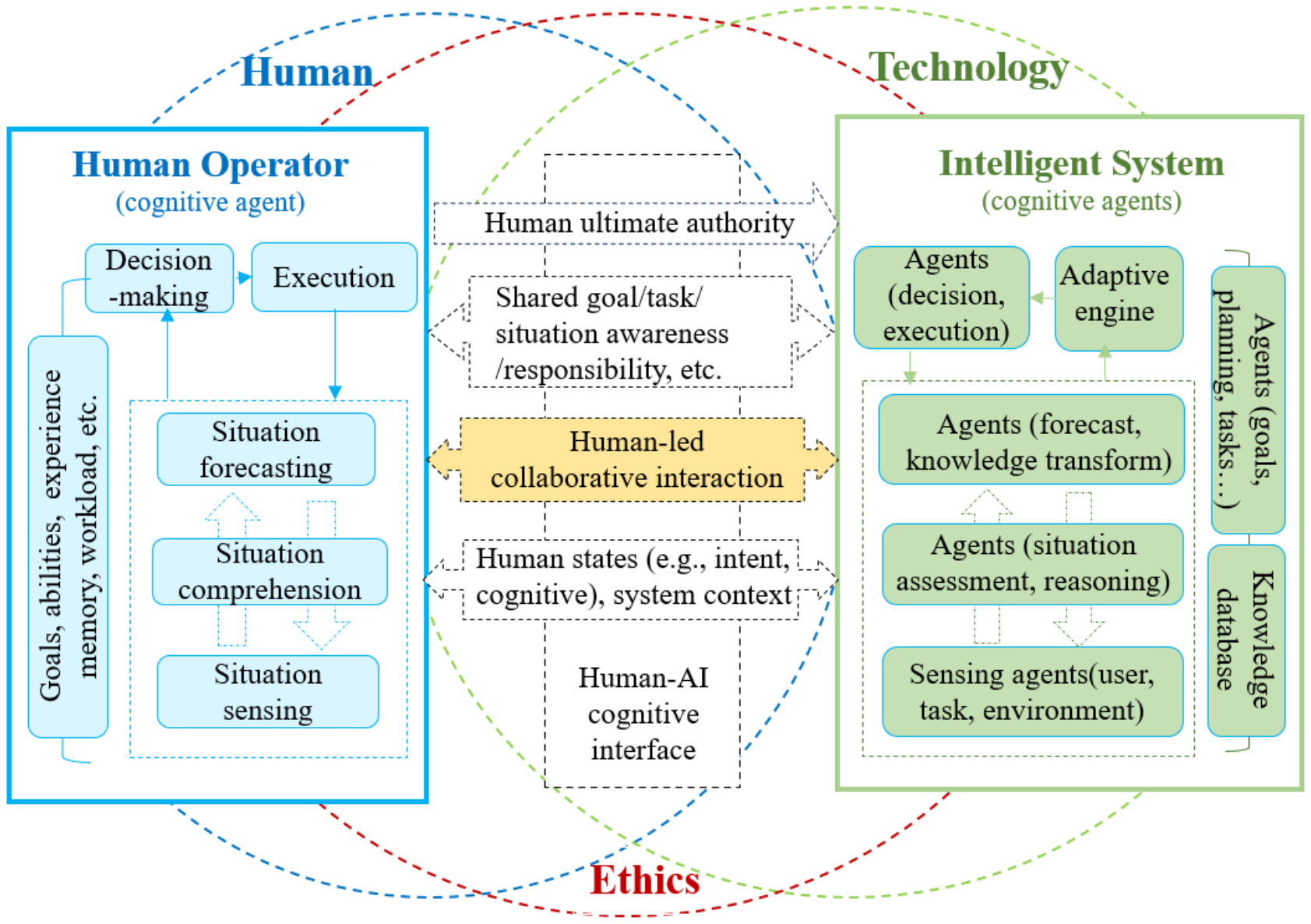

**Fig. 6** A conceptual framework of human-AI joint cognitive systems (adapted from Xu et al, [121])

disciplines through training, tools, and open dialogue.

- *Measure success holistically:* Evaluate outcomes using metrics like user satisfaction, and ethical impact.

*E. HCAI Multi-Level Design Paradigms*

A design paradigm provides a specific perspective that shapes the scope, approaches, and methods within a field. In the computer era, different design paradigms were used to tackle challenges. For instance, human factors science applies the paradigm of human cognitive information processing, studying how psychological activities (e.g., perception and attention, interact with systems [119]. Fields like HCI and UX treat computers as supportive tools, focusing on user models, UI concepts, and usability testing for good experience [120].

As an emerging field, HCAI needs effective design paradigms to tackle new challenges. Recent research highlights the impact of various factors on HCAI practice and the need for effective paradigms to address limitations in HCAI practices [11], [37]. This paper extends our previous work to align with the HCAI multi-level design paradigms (HCAI-MDP) as a component of HCAI-MF [121], [122]. This extension offers new design thinking for HCAI practice.

**1) Human-AI joint cognitive systems**

Inspired by the frameworks of joint cognitive systems and situation awareness [123]. [124], Xu et al. [121] introduced a conceptual framework of human-AI joint cognitive systems, representing the human-AI collaborative interaction (Figure 6). Unlike traditional HCI models, the framework treats an AI system (e.g., an autonomous vehicle) as a cognitive agent (or agents) capable of performing specific cognitive information-processing tasks. This framework views a human-AI system consisting of two cognitive agents: human and AI. The AI agent engages in two-way collaborative interactions with users via natural interfaces (e.g., voice input) and performs tasks like learning, recognizing user states (e.g., intentions, emotion) making decisions, and executing actions guided by humans.

The framework uses Endsley's theory of situation awareness on how human and AI agents process information [124]. Humans perceive and interpret their environment, predict future states, and rely on memory, experience, and knowledge. Similarly, AI agents autonomously sense, reason, decide, and act, extending human cognitive principles to AI systems.

While current AI systems are not fully collaborative, viewing them as cognitive agents encourages advancements in AI technology and HCAI systems [11]. This framework emphasizes collaborative interaction, with humans taking the lead, unlike traditional approaches that separate human and AI capabilities. Enabling collaborative interaction establishes a foundation for HCAI systems and introduces a new design paradigm for integrating human and AI strengths. This framework presents a new design paradigm, as outlined below:

*Machine-Based Cognitive Agents:* Unlike traditional HCI design paradigms, this paradigm shifts from viewing machines as tools to treating them as potential collaborators. It enhances human-AI system performance by analyzing and optimizing the collaborative interactions of both human and AI agents, fostering a symbiotic relationship for maximum performance.

*Human-Centered Approach:* This paradigm ensures humans lead and control human-AI systems, especially in critical situations. It integrates human factors, technology, and ethics perspectives (represented by the overlapping zone of the three circles in Figure 6 [1]) to align with HCAI principles, prioritizing humans while leveraging advanced AI technology.

*Human-Led Collaborative Interaction with AI:* This framework redefines the human-machine relationship as a joint cognitive system with two cognitive agents, moving beyond traditional "interaction" models. Inspired by the theories of human-human teamwork, it provides a basis for studying, modeling, designing, building, and validating collaborative interaction, enhancing overall system performance.

*Bidirectional Human-AI Interaction:* Unlike the traditional one-way "stimulus-response" model [125], this paradigm emphasizes two-way interaction. AI systems monitor users' physiological, emotional, and intentional states using advanced sensing technologies, while humans gain situational awareness through multimodal user interfaces. This dynamic interaction supports collaboration-based interfaces, improving human-AI interaction by addressing key factors like shared situation awareness, mutual trust, and human-led joint decision-making.

The design paradigm of human-AI joint cognitive systems is aligned with the HCAI guiding principles (Table 2). For example, treating AI as a cognitive agent promotes *Transparency and Explainability* through two-way, situation-aware interactions that help users understand AI decisions. It emphasizes *Human Control & Empowerment* to ensure human authority in critical contexts. The paradigm integrates ethical, technological, and human perspectives, directly supporting *Ethical Alignment*. Its design enhances *User Experience* with natural, multimodal interaction. It focuses on *Human Augmentation & Human-Led Collaborative Interaction with AI* by optimizing the strengths of both human and AI agents, enhancing system productivity. Lastly, by ensuring resilient, situation-aware behavior of AI agents, it addresses *Safety & Robustness. Accountability* is reinforced by maintaining human oversight and responsibility over AI behaviors.

In reality, full situation awareness and real-time adaptation remain challenging. This paradigm is scalable for gradual adoption based on technology maturity and domain needs through three strategies: (1) *Multi-level design:* integrated with the HCAI approach for partial implementation, where full adaptation is not possible. (2) *Human-in/on-the-loop control*: bidirectional interactions and modular role allocation enable human-led oversight and control, even without full real-time awareness, such as driver's manual overrides to balance autonomy with supervision. (3) *Adaptation without full autonomy*: designed for hybrid intelligence and shared operations. In low-adaptivity settings, AI offers support instead of acting autonomously. Thus, the paradigm does step-by-step integration as an evolving solution, not all-or-nothing approach.

**2) Human-AI joint cognitive ecosystems**

The human-AI joint cognitive systems framework focuses on individual AI systems. As AI evolves, intelligent ecosystems such as smart cities and healthcare emerge, involving interactions among humans, AI systems, and AI-to-AI systems. The safety and performance of these ecosystems depend on effective synergy across subsystems.

For example, human-vehicle co-driving in automated vehicles (AVs) represents a single human-AI joint cognitive system. An AV ecosystem, however, includes collaboration between drivers and onboard AI in individual AVs, interactions among AVs, and integration with intelligent traffic and transportation systems. The safety and performance of the AV ecosystem depend on optimizing individual human-AV systems and coordination across all ecosystem components.

Research on multiple AI systems in intelligent ecosystems has mainly focused on technology with limited attention to human-AI collaborative interaction [126], [127]. Inspired by joint cognitive systems theory [105], multi-agent systems theory [128], and ecosystem theories [127], [129], Xu et al., [121] proposed a conceptual framework of human-AI joint cognitive ecosystems. These ecosystems consist of interconnected human-AI systems, human-human interactions, and AI-AI interactions (Figure 7). The human-AI joint cognitive ecosystem framework shifts HCAI focus from individual AI systems to the broader ecosystem, offering a transformative perspective for HCAI practices.

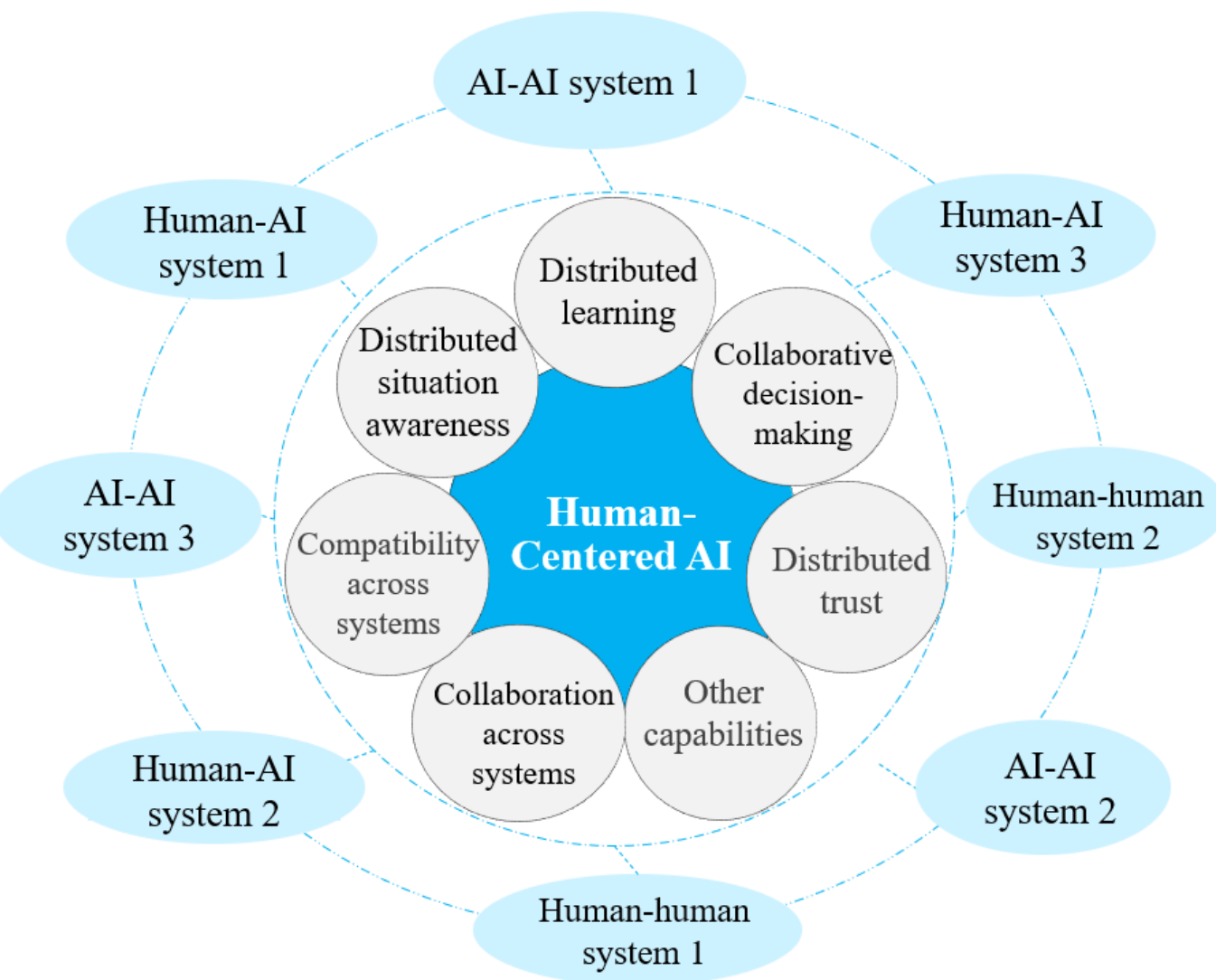

**Fig. 7** A conceptual framework of human-AI joint cognitive ecosystems (adapted from Xu et al [121])

*Systematic Design Thinking:* Unlike the human-AI joint cognitive systems framework focusing on individual human-AI systems, this paradigm emphasizes collaborative interaction among multiple human-AI systems, human-human systems, and AI-AI systems within an ecosystem. It advocates a systematic approach to optimizing the design of intelligent ecosystems by addressing the interplay among all subsystems. It shifts the focus from a "point" design (individual human-AI systems) to a "2-D plane" design, considering the collective dynamics and collaboration across the ecosystem.

*Human-Centered Approach:* This design paradigm takes a human-centered from an ecosystem-wide perspective, ensuring humans remain central to all aspects of the ecosystem. Examples include building a human-AI/AI-AI/human-human trust ecosystem across ecosystems, resolving conflicts between AI systems from different vendors with varying ethical standards, and creating security frameworks for decision-making authority, ensuring humans maintain ultimate control.

*Distributed Collaborative interaction:* Given the inherently distributed nature of intelligent ecosystems, their design must address a range of collaborative interactions characterized by distributed and shared responsibilities. Key elements include shared situation awareness, human-AI/AI-AI/human-human co-learning, and joint decision-making with human authority, enhancing the collaborative interaction across ecosystems.

*Ecological Learning and Evolution:* Like natural ecosystems, intelligent ecosystems evolve based on the learning and adaptation of human and AI agents. This paradigm emphasizes continuous learning, optimization, and adaptation through mechanisms like cross-task knowledge transfer, self-organization, and adaptive collaboration. These capabilities allow the ecosystem to adapt to dynamic scenarios.

In summary, this design paradigm aligns with the HCAI guiding principles. By extending the focus to ecosystems, it emphasizes *Transparency and Explainability* through shared situation awareness and ecosystem-level decision-making visibility. It reinforces *User Control and Empowerment* to ensure human authority, even in multi-agent settings. It explicitly addresses *Ethical Alignment* by accounting for value conflicts and promoting harmonized ethical behavior within an ecosystem. It improves *User Experience* through systematic design that integrates human-human, human-AI, and AI-AI interactions. *Human Augmentation & Collaborative Interaction with AI* fosters symbiotic, distributed collaboration to enhance both individual and collective performance. It prioritizes *Safety & Robustness* for coordinated interactions, and adaptive behaviors across subsystems. *Accountability* is preserved through human oversight and shared responsibilities, while *Sustainability* is supported through ecological learning that drive long-term adaptability, self-organization, and alignment with broader human and societal needs.

**3) Intelligent sociotechnical systems**

AI systems and intelligent ecosystems operate within a broader sociotechnical system (STS) environment. According to STS theory, achieving optimal performance requires the joint optimization of society, technology, organizations, and other subsystems [130]. This principle applies to both individual human-AI systems and intelligent ecosystems.

AI introduces unique challenges, such as impacts on privacy, ethics, decision-making, reskilling, and work systems (e.g., role, flow). These require designing and deploying AI within a macro sociotechnical context [83]. However, there is no established STS theory tailored to AI technologies. To address this gap, Xu et al. [122] proposed a conceptual framework of intelligent sociotechnical systems (iSTS) (Figure 8).

Figure 8 shows that iSTS retains key features of traditional STS theory, including interdependent technical and social subsystems, with overall performance relying on their joint optimization [130], [131]. Unlike the human-AI joint cognitive ecosystem, iSTS emphasizes macro and non-technical factors' influence on AI for HCAI practice, viewing AI system design as both an engineering and sociotechnical challenge that demands interdisciplinary collaboration and a systematic approach. It offers a new design paradigm for HCAI practices.

*Systematic Design Thinking:* The development of human-AI systems or intelligent ecosystems takes place within a sociotechnical environment. The iSTS design paradigm provides a holistic approach to optimizing these systems by integrating both technical and non-technical subsystems, moving beyond purely technical considerations.

*Human-Centered Approach:* This paradigm highlights the impacts of non-technical factors like humans, society, culture, and ethics on the design and use of AI systems. By incorporating these factors, it ensures alignment with HCAI principles, enabling AI technology to support informed decision-making, augment human capabilities, and maintain human authority as the ultimate decision-maker.

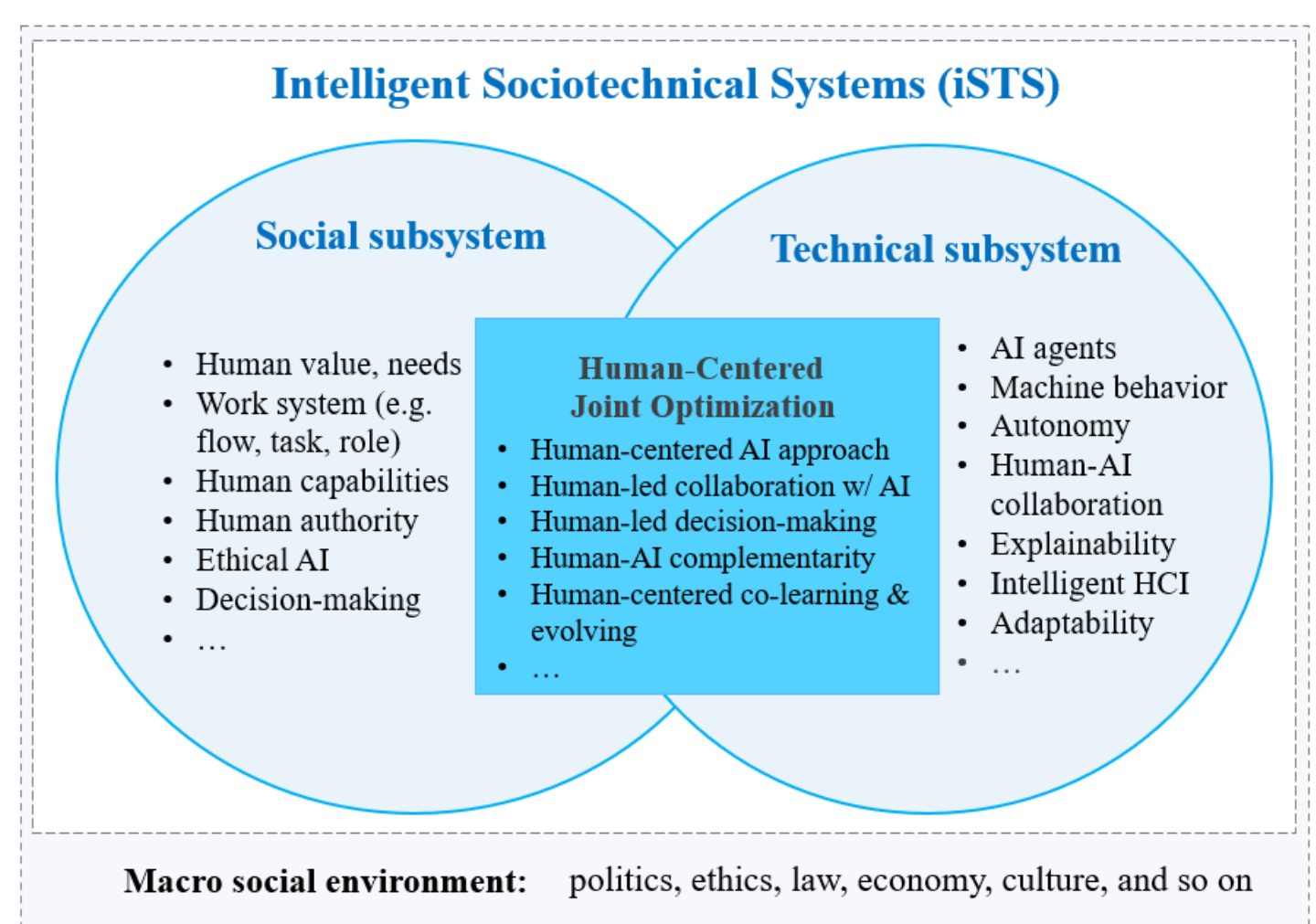

**Fig. 8** A conceptual framework of intelligent sociotechnical systems (iSTS) (adapted from Xu & Gao [122])

*Human-AI Collaborative Interaction in a Macro Context:* The overlapping region of the AI technology and social subsystem in Figure 8 highlights the synergistic integration between the technical and social subsystems. Unlike traditional STS theory that views technologies as tools to boost productivity, iSTS envisions AI as potential collaborators, working alongside humans to build trust and enable shared decision-making with human authority, enhancing deeper human-AI collaborative interaction and overall performance.

*Work System Redesign:* AI technology will reshape existing work systems, creating challenges in balancing machine autonomy with human control. iSTS emphasizes redesigning tasks, roles, and workflows to leverage the complementary strengths of humans and AI. This redesign prioritizes fairness, user satisfaction, participatory decision-making, and skill development, aiming to enhance overall performance.

*Human-AI Co-Learning and Evolution:* In an iSTS environment, humans and AI interact across individual, organizational, and societal levels, enabling mutual adaptation. AI agents act as a novel resource that facilitates coordination between the social and technical subsystems. AI learns through algorithms, while humans adjust expectations through social learning. Both adapt and evolve together, highlighting the need to foster co-learning, co-evolving, and mutual growth [132].

*Open Ecosystem:* Unlike traditional STS focusing on defined organizational boundaries, iSTS functions within dynamic and interconnected ecosystems. The rise of autonomous AI agents increases uncertainty, dynamism, and fuzzy boundaries of iSTS. This requires a shift to an open ecosystem perspective to manage their evolving and unpredictable nature [133].

To sum up, the iSTS paradigm advances HCAI guiding principles by systematically integrating technical and social subsystems within a broader societal context. It promotes *Transparency and Explainability* by fostering human-AI collaborative interaction to support trust and decision-making. It reinforces *User Control & Empowerment* to ensure human's ultimate decision-makers within evolving ecosystems. iSTS emphasizes *Ethical Alignment* by incorporating cultural, societal, and ethical factors into AI systems. It enhances *User Experience* through redesigned work systems that prioritize human needs. It advances *Human Augmentation & Human-Led Collaboration with AI* to optimize complementary strengths. It addresses *Safety & Robustness* through interdisciplinary design to enhance system resilience. *Accountability* is maintained through systematic human oversight in sociotechnical operations. Lastly, its open, adaptive, and evolutionary approach supports *Sustainability* to promote long-term human-AI co-learning, responsible, ecosystem-wide development.

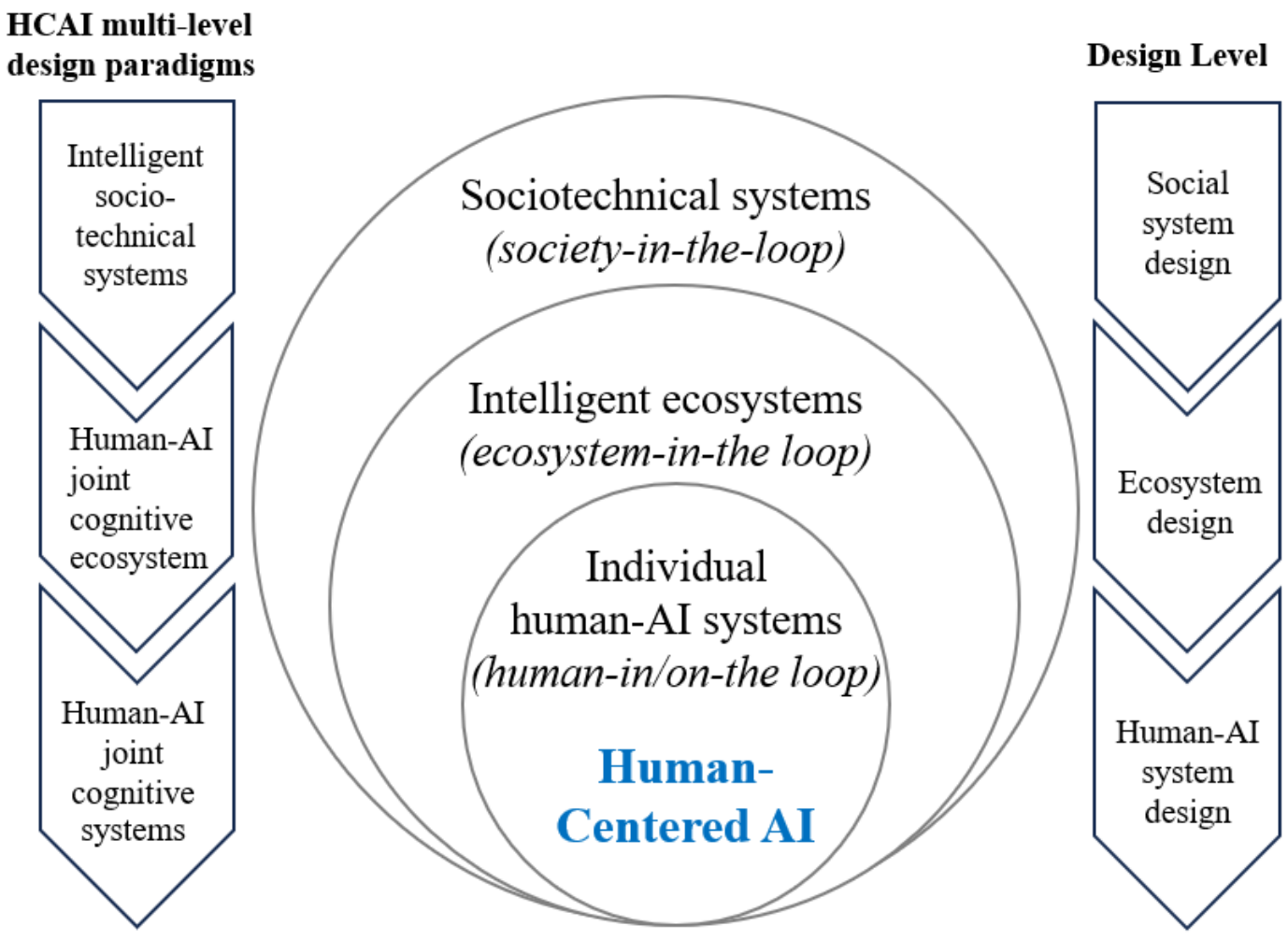

**Fig.9** Illustration of HCAI multi-level design paradigms

Figure 9 illustrates our proposal that HCAI should be practiced from ecosystems and sociotechnical perspectives, beyond the current focus on individual human-AI systems. This paradigmatic shift in HCAI practice addresses existing limitations with a comprehensive approach [40], [122]. The multi-level design paradigms offer benefits to HCAI practice:

• *Systematic design thinking*: Introduces a systematic strategy to tackle AI challenges with a multi-level approach

• *An expanded scope*: Shifts the focus beyond individual human-AI systems to ecosystems and sociotechnical contexts.

• *Strategic planning*: Provides a framework for defining HCAI practical strategies early in AI development.

• *Human-centered approach*: Integrates HCAI approach across multiple design paradigms and broader contexts.

Table 3 further summarizes the significance of the three design paradigms for HCAI practice. Among them, the column headings in Table 3 represent the diversified paradigms that can support the key HCAI topics listed in Column 1. These paradigms also include existing design paradigms. The cells in Table 3 summarize the critical issues to be addressed across these key HCAI topics by leveraging these design paradigms.

As highlighted by Table 3, future HCAI practice needs diversified design paradigms; the existing design paradigms still play an essential role, and the three emerging design paradigms play a supplementary role. The three emerging design paradigms help broaden HCAI practice to effectively address the new challenges of AI for creating HCAI systems.

## V. Case Study

### A. Human-Centered Co-driving in Automated Vehicles

AI-based automated vehicles (AVs) exemplify a typical human-AI system. SAE classifies AV automation into five levels [134]. According to SAE, AVs at or below Level 4 still require human drivers to participate in driving. Consequently, human-AV co-driving is expected to remain the norm for a significant period. Despite the involvement of human factors professionals, public trust in AVs remains low [135]. Frequent AV-related accidents have highlighted the need for new design approaches to improve safety and trust [136].

From the HCAI perspective, human-AV co-driving at SAE Levels 3 and 4 presents challenges in ensuring safety, trust, and collaboration between drivers and AVs. Previous research has identified challenges. For example, drivers often become distracted or disengaged during automated driving, making it challenging to resume control when requested [137], [138]. Poorly designed handover mechanisms and human-AV interfaces can lead to confusion about who is in control, increasing risk and delayed reactions [139], [140]. Over-trust may cause drivers to become complacent, while under-trust can lead to unsafe overrides [141]. Current practices also have limitations. For example, design is focused on individual human-AV systems and the technical aspects of the co-driving system. Such a narrow focus does not fully consider the impacts of the AV ecosystem and non-technical aspects from the broader sociotechnical perspective, such as coordination across ecosystem components, human ethics, public awareness and acceptance, and driver behaviors [40], [142], [143], [144]. As a result, such an isolated approach cannot guarantee the optimized design and safety of the entire AV ecosystem.

To address these challenges and the narrow focus in current research and design, we conducted a gap and needs analysis by applying HCAI-MF. The goal was to analyze HCAI-related challenges in human-AV co-driving and provide insights to inform the requirements and design for human-centered co-driving solutions. Gap and needs analysis, as initial analysis in the early stage, are essential in defining requirements to ensure that AI systems are aligned with design goals. It helps identify mismatches between the current state of system capabilities and the desired outcomes, driving requirement definitions. The analysis was informed by multiple sources reflecting diverse stakeholder perspectives, including a literature review and analysis (see key references in Appendix C) and insights from industry experts in the AV domain obtained through interviews and joint working sessions.

### B. An HCAI Approach

#### 1) Gap and needs analysis based on applying HCAI design paradigms

Our gap and needs analysis started with the HCAI multi-level design paradigms with the following assumptions.

• *Human-AI Joint Cognitive Systems:* In-vehicle AI systems equipped with sensing/control technologies can perform tasks like sensing, recognition, and inference based on the driver's state and environment. Drivers and the AI agents work as two cognitive agents, complementing each other. The co-driving can be represented by a human-AI joint cognitive system.

Table 3 Implications of design paradigms for HCAI key research topics

| Research items | Computational approach (AI/CS) | Human-centered design (human factors science) | Human-AI joint cognitive systems | Human-AI joint cognitive ecosystems | Intelligent sociotechnical systems |
|---|---|---|---|---|---|
| AI-based machine behavior | Algorithm optimization to minimize system bias. | Apply user-participatory testing methods in algorithm training and tuning to minimize bias. | The impact of human-AI collaboration on machine behavior (co-learning, co-evolving). | Machine learning and behavior evolution across multiple AI systems within human-AI ecosystems. | Effects of the social factors on machine behavior, ethics of machine behavior. |
| Human-AI collaboration | Computational models of recognizing user states (e.g., intent). | Collaboration-based cognitive interface model, design, and user validation. | Human-AI collaboration theory (mutual trust, shared control, and human leadership). | Collaboration, adaptation, co-learning, and evolution mechanisms across multiple AI systems. | Social impacts of human-AI collaboration, social responsibility on collaboration. |
| Human-machine hybrid intelligence | Advanced human cognition modeling, “data + knowledge” dual-driven AI. | The “human-in/on-the-loop” interaction design; complementary design of human-AI advantages. | Models and methods of human-AI hybrid intelligence based on human-AI collaboration. | Distributed collaborative cognition, human-AI co-learning theory across multiple AI systems. | Human-AI function allocation, authority setting in the social environment. |
| Ethical AI | Algorithm governance, algorithm training & optimization, reusable ethical AI codes. | Controllable AI design, meaningful human control, behavioral methods for ethical AI design. | Ethical issues of human-AI collaboration (e.g., norms, authority, culture). | Human authority setting and compatibility across multiple AI systems due to different cultures/norms, | Ethical AI issues in the social environment, ethical AI norms and governance. |
| Human-AI interaction | Algorithms for intelligent interaction (e.g., emotional, intent). | New UI paradigms for intelligent systems, intelligent interaction design standards. | Collaboration-based cognitive UI, interaction design paradigms based on human-AI teaming. | Interaction compatibility theory, technology, and design across multiple AI systems. | Impacts of the social environment (cultural, ethical, etc.) on interaction design. |

- *Human-AI Joint Cognitive Ecosystems:* The safety and efficiency of human-AV co-driving rely not only on the driver-AV interaction design of individual AV systems but also on collaboration among various agents, such as drivers, AVs, pedestrians, intelligent road systems, and transportation infrastructure. These collaborations form a human-AI joint cognitive ecosystem, emphasizing interdependence and joint optimization design within the entire co-driving ecosystem.
- *Intelligent Sociotechnical Systems (iSTS):* Human-AV co-driving is both a technical and sociotechnical project. Achieving safety and efficiency requires optimizing AV technical elements alongside non-technical factors like ethical standards, AV traffic infrastructure, regulations, public trust, driver behavior and skills, stakeholder collaboration (e.g., drivers, AV makers, policymakers), and AV certification. The iSTS paradigm integrates diverse technical and non-technical factors to create safe, reliable, and accepted AV systems.

Based on a literature review, we identified a list of gaps and challenges in the current approaches to human-AV co-driving. We then proposed the high-level needs to address these challenges and gaps based on the three HCAI design paradigms. Appendix C outlines the results of the gap and needs analysis.

The gap and needs analysis grounded in the three HCAI design paradigms, each embedding core HCAI guiding principles, highlights key shortcomings in current AV design approaches and provides actionable insights for developing human-centered co-driving systems.

At the *human-AI joint cognitive systems* level, prevailing AV designs are technology-centric, often treating AVs as tools rather than collaborative agents. This results in unclear driver roles, unidirectional interaction models, and inadequate human-AI complementarity. To address the gaps, the analysis recommends adopting an HCAI approach where AVs are collaborators, promoting dynamic role sharing, adaptive and collaborative user interfaces, and meaningful human control.

At the *human-AI joint cognitive ecosystems* level, current systems focus too narrowly on individual driver-AV systems, neglecting broader interactions with pedestrians, infrastructure, and other AVs. This limited scope overlooks the importance of distributed collaboration, ecosystem-level safety, and adaptive learning among all agents. The solution lies in designing human-AI joint cognitive ecosystems where all human and AI agents collaborate through shared situational awareness, trust, control, and evolving coordination mechanisms.

Finally, from the perspective of *intelligent sociotechnical systems*, current designs are insufficiently addressing issues such as social integration, ethical dilemmas, and public trust. Non-technical factors such as human values, policy coordination, user data privacy, and behavioral adaptation are often ignored. The analysis highlights the needs that optimize both technical and social subsystems, enforce harmonized ethical and safety standards, ensure equitable access and trust, and facilitate cross-sector collaboration among stakeholders.

Figure 10 further illustrates the implications of these three HCAI design paradigms, expanding the co-driving solution from individual human-AI systems (a "point" solution) to AV ecosystems across multiple human-AI/ human-human/AI-AI systems within (a "2-D plane" solution) and finally to sociotechnical systems integrating technical and non-technical subsystems (a "3-D cube" solution). Such a comprehensive approach goes beyond isolated strategies, setting new requirements for a human-AV co-driving solution. Implementation strategies can proceed along three pathways:

- *A bottom-up approach*: Start with individual human-AV systems, expand to AV ecosystems and sociotechnical systems.
- *A top-down approach:* Begin with a strategic vision of the AV sociotechnical systems, then focus on the AV ecosystem, and finally address individual human-AV systems.
- *A hybrid approach:* Combine bottom-up and top-down strategies to balance both strengths for more effective solutions.

**2) Mapping analysis of other HCAI-MF components**

The HCAI design paradigms drove the perspective and strategy of the work. Following the gap and needs analysis, we defined the high-level requirements of human-AV co-driving by leveraging the other HCAI-MF components. This process was also facilitated by the HCAI process and HCAI interdisciplinary collaboration involving experts across domains such as AV, system engineering, and human factors. Appendix D presents high-level requirements for human-AI co-driving by mapping HCAI design paradigms, guiding principles, and methods from the perspective of the Human-AI Joint Cognitive Systems paradigm.

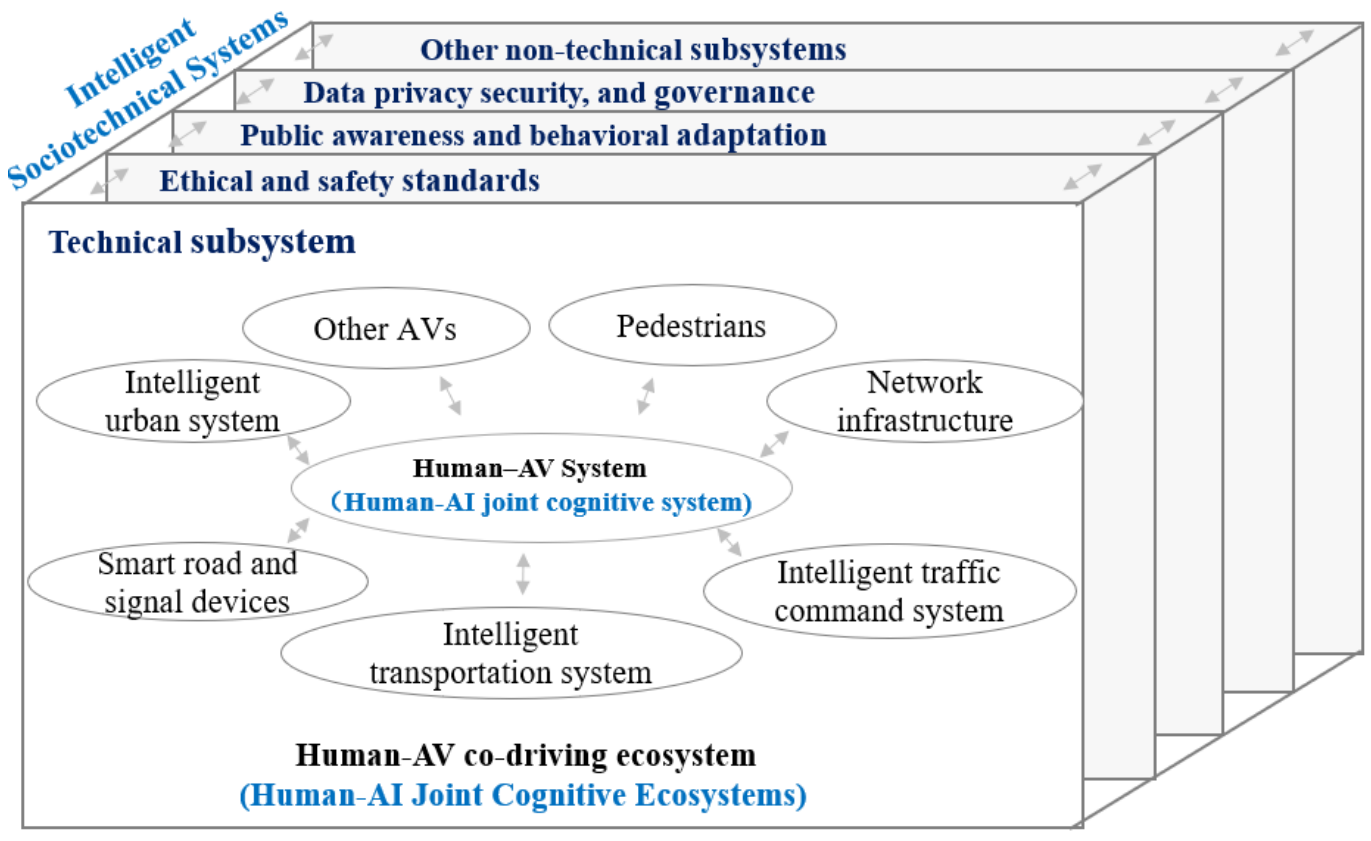

**Fig. 10** Illustration of applying HCAI multi-level design paradigms to human-AV co-driving

The mapping analysis was carried out through the steps outlined from left to right in Appendix D:

- Defined the design themes per HCAI design paradigms
- Defined the key design actions guided by HCAI principles
- Determined the HCAI approaches & methods applied
- Defined HCAI-based high-level requirements

As an illustration of the analysis, grounded in the paradigm of Human-AI Joint Cognitive Systems, this mapping analysis generates a set of HCAI-driven high-level requirements for an HCAI-based co-driving solution. The requirements emphasize a collaborative and dynamic allocation of roles between the driver and AV. The system should be designed to enhance human control, support intuitive user experiences, and promote transparency and explainability of AV behaviors through multimodal interfaces. It calls for human-led collaboration, where AVs augment rather than replace humans, ensuring meaningful human oversight via "human-in/on-the-loop" designs. Critical functions include shared situation awareness and collaborative decision-making, enabling drivers to accept, reject, or override AV suggestions. The results further recommend driver state monitoring to adapt AV behavior in real time, enhancing safety and personalization. Adaptive and collaborative interaction should ensure fluid transitions of control for AVs to take charge when drivers are unresponsive. These requirements aim to foster a human-centered, collaborative, and context-aware co-driving experience.

This analysis also illustrates the value of the HCAI requirement hierarchy (Figure 3) and the HCAI approach and method taxonomy (Appendix A), which provides a structured approach for deriving HCAI requirements and actionable design insights for the human-AV co-driving solution.

*C. Summary*

Collectively, the gap and needs analysis and the mapping analysis both underscore the necessity of reimagining AVs not as isolated technologies but as integrated, adaptive, and trustworthy components of human-centered sociotechnical ecosystems; the work provides a holistic view of human-AV co-driving systems, addressing both technical and non-technical aspects. As for the next steps, we are currently engaging additional stakeholders to validate these preliminary findings and to develop more detailed requirements. This case study shows how HCAI-MF can foster strategic HCAI design thinking and generate actionable insights, providing a foundation to address challenges in current human-AV co-driving practices, with the goal of developing a safe, human-centered co-driving solution.

## V. Discussions

### A. Contributions of HCAI-MF

#### 1) Addressing Gaps in Current HCAI Practice

HCAI-MF overcomes the gaps in current HCAI practice with a comprehensive methodological framework, addressing specific challenges through its components: (1) HCAI requirement hierarchy: Provides a structured approach to improve HCAI requirement gathering. (2) HCAI approach and method taxonomy: Categorizes actionable HCAI approaches and methods to operationalize HCAI design goals. (3) HCAI process: Defines systematic workflows to resolve limitations in current processes. (4) HCAI interdisciplinary collaboration approach: Tackles challenges in cross-disciplinary teamwork essential for HCAI. (5) HCAI multi-level design paradigms: Expands the scope of HCAI practice to include joint cognitive systems, ecosystems, and sociotechnical perspectives.

#### 2) An Executable HCAI Methodology

HCAI-MF transforms HCAI from a design philosophy into a methodological framework. With the five key components, it provides actionable guidance, practical processes, and design paradigms to implement HCAI across the AI system lifecycle, helping organizations deliver human-centered AI systems.

#### 3) A Structured Approach for Flexibility and Scalability

HCAI-MF provides a structured approach to guide HCAI practices, ensuring a consistent approach while allowing flexibility for customization. Project teams can adapt elements in the HCAI requirement hierarchy and process for specific needs. Its scalable approach supports ongoing updates, such as refining guiding principles or introducing new methods, ensuring HCAI-MF stays relevant as HCAI practice evolves.

#### 4) A future research agenda

While HCAI-MF proposes actionable guidance and requirements to close the gap in current HCAI practice, it also provides a work agenda for future HCAI research from a methodological perspective. HCAI methodology must be continually enhanced through ongoing research and refinement of HCAI approaches, supported by strong interdisciplinary collaboration that integrates diverse approaches, methods, and paradigms. HCAI-MF provides a structured perspective towards a mature HCAI methodology. Future work should continue to advance HCAI methodology and its components.

### B. Challenges and Recommendations for Executing HCAI-MF

Like the early days of HCD in the 1980s, implementing

Table 4 The potential challenges in implementing HCAI-MF and recommendations to address the challenges

| Areas | Potential challenges in practice | Actionable recommendations for practicing HCAI-MF within organizations |
|---|---|---|
| Overall HCAI-MF implementation strategy [23], [159] | Difficulty in initiating HCAI-MF | Tailor HCAI-MF to fit different sizes of organizations, determine based on maturity and priority. |
| | Misalignment with business objectives | Align HCAI with organizational goals, demonstrate value with pilot projects/measurable results. |
| | Change resistance in culture | Promote leadership advocacy of HCAI, show successful projects, and offer employee training |
| | Challenges in measuring success | Develop key performance indicators (KPI) to track trust, user experience, and ethical compliance. |
| | Scalability of HCAI-MF practices | Create standardized toolkits, templates, processes, training for consistent/scalable HCAI practice. |
| | Resource constraints (time/budget/skill) | Prioritize and fund high-impact projects for HCAI initiatives, leverage open-source AI tools. |
| HCAI requirement hierarchy [26], [34] | *Recommended steps (Figure 3)*: (1) Select or develop HCAI design guiding principles; (2) Define or adopt specific HCAI design guidelines to support the HCAI guiding principles; (3) Define product detail requirements supported by human-centered requirements gathering methods. | |
| | Lack of understanding of design principles | Conduct training, create organization-specific guidelines, use case studies illustrating principles. |
| | Overgeneralized design guidelines | Tailor guidelines to specific applications or use cases, ensuring they are actionable/relevant. |
| | Resistance in adopting HCAI principles | Use change management, showcase benefits through initiatives, gain leadership support early. |
| | Difficulty in operationalizing principles | Translate principles into actionable guidelines, create checklists, and integrate into workflows. |
| HCAI approach & method taxonomy [5], [64] | *Recommended steps (Appendix A)*: (1) Determine the human factors (e.g., human controllability) to be prioritized with appropriate HCAI design principles (See Table 2); (2) Choose appropriate approaches and methods that support the human factors to be prioritized. | |
| | Lack of expertise in HCAI design | Provide training, hire HCAI experts, and foster cross-disciplinary collaboration. |
| | Complexity in integrating HCAI approach | Create a phased plan, integrate HCAI into current workflows, standardize approaches & methods. |
| | Conflicting methodologies | Encourage flexibility and compromise and select methodologies to align with HCAI. |
| | Insufficient user involvement | Engage users at every design stage, leverage diverse users, and use a participatory design method. |
| HCAI process [8], [28] | *Recommended steps (Figure 5)*: (1) Define a set of HCAI activities for projects; (2) Integrate the HCAI activities into the existing AI system lifecycle in terms of milestones, deliverables, and resources; (3) Track the progress of HCAI activities and take corrective actions as needed. | |
| | Difficulty in integrating workflows | Develop standardized processes/tools for seamless integration of contributions across disciplines. |
| | High costs and resource requirements | Start with smaller, impactful projects, secure executive sponsorship, leverage open-source tools. |
| | Varying levels of commitment | Clarify roles and responsibilities and provide incentives/recognition for participation in projects. |
| | Limited expertise in HCAI design | Provide training, hire experts, foster cross-functional collaboration to build internal capabilities. |
| HCAI interdisciplinary collaboration approach [8], [38], [39], [159] | *Recommended steps (Section IV.D)*: (1) Set up an interdisciplinary team including HCAI-related experts; (2) Define communication channels and decision-making process; manage conflicts in ethical standards or regulatory interpretations; (3) Define disciplinary activities and deliverables in the project plan. | |
| | Differing terminologies and perspectives | Facilitate training for a shared vocabulary and understanding across functional teams/disciplines. |
| | Resistance to collaboration | Promote team-building activities, open communication, and a culture of mutual respect. |
| | Fragmented team structure | Organize teams around defined projects/goals and appoint a manager for interdisciplinary efforts. |
| | Conflicts in ethical standards x-disciplines | Set up a shared baseline & interdisciplinary boards for universal principles x-cultures /disciplines |

HCAI-MF faces challenges, as HCAI is still unfamiliar to many. For example, poor interdisciplinary collaboration, with human-centered professionals lacking AI knowledge and AI professionals struggling to understand other fields [5], [37]. This disconnect creates collaboration difficulties, with non-AI experts reporting misunderstandings due to the lack of shared workflows and common language [158]. Without effective strategies, the potential of HCAI-MF cannot be fully realized.

Based on lessons learned from HCD practice and literature review, Table 4 outlines potential challenges in implementing HCAI-MF and offers the overall strategy and actionable recommendations for HCAI-MF implementation.

To support HCAI-MF adoption, a "three-layer" HCAI-MF implementation strategy is proposed across three levels as supported by available approaches and tools (Figure 11): (1) *Project teams* focus on practical applications in individual projects by implementing the proposed HCAI-MF and currently available HCAI practical resources [4], [8], [26], [28], [29], [160]; (2) *Organizations* embed HCAI culture into organizational environments and processes by adopting HCAI enterprise strategies [159], organization redesign [162], business process redesign [163], and HCAI maturity model [164]. Our proposed HCAI maturity model (HCAI-MM) can work as a diagnostic tool to assess an organization's current HCAI maturity, identify gaps, and create actionable roadmaps [164]; (3) *Societal-level efforts* require an HCAI-driven societal transformation [165] and promote support for HCAI by implementing international guidelines and frameworks [30], [52], [116], [155]. This strategy ensures effective and sustainable implementation of HCAI-MF.

### C. Future Work of HCAI-MF

#### 1) Enhancing HCAI-MF

HCAI-MF must evolve to support future practices by updating HCAI approaches and methods through ongoing research and practical refinement. Robust metrics and evaluation tools are essential to measure its effectiveness. Additionally, tools like HCAI maturity model should be developed to facilitate HCAI-MF implementation.

#### 2) Practicing HCAI-MF Across Domains

Applying HCAI-MF across various domains, such as healthcare, transportation, education, LLMs, extended reality, and human-robot interaction present challenges that need further exploration. Conducting domain-specific case studies is crucial to evaluating their effectiveness in real-world settings. These studies can validate HCAI-MF, identify challenges, and uncover opportunities for improvement, ensuring HCAI-MF remains relevant and adaptable across diverse applications.

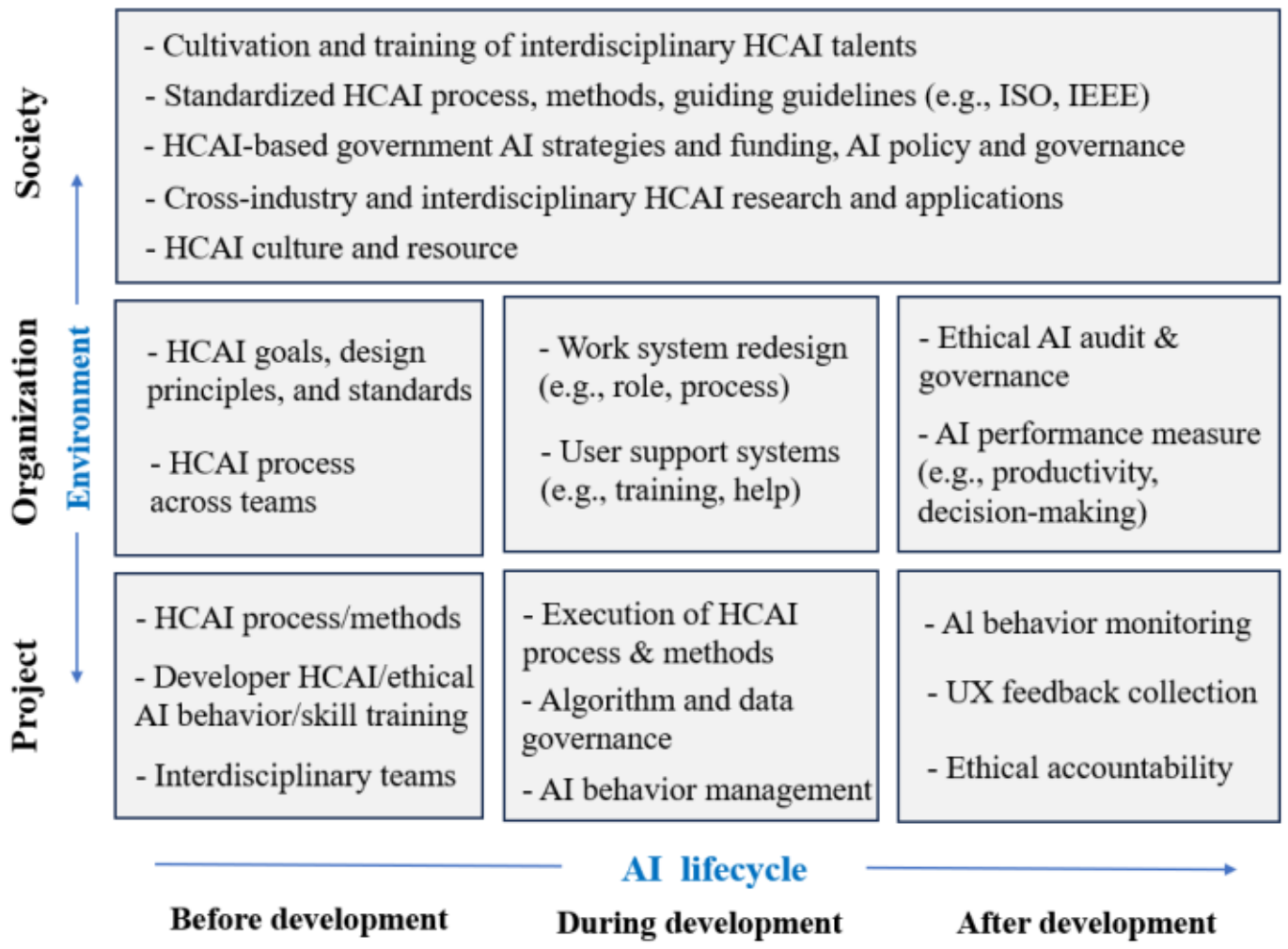

**Fig. 11** A "three-layer" HCAI-MF implementation strategy

**3) Fostering interdisciplinary collaboration**

The successful implementation of HCAI-MF depends on strong interdisciplinary collaboration across approaches, methods, processes, and design paradigms. A long-term strategy should focus on developing interdisciplinary HCAI talent, like how HCI and UX expertise have evolved. Universities could offer programs like "AI Major + Minor" or "Major + AI Minor" to combine AI with complementary fields. Government initiatives should promote HCAI design and fund interdisciplinary research, while cross-industry projects can foster collaboration between academia and industry.

## VI. CONCLUSIONS

HCAI, as a design philosophy, has influenced the design, development, deployment, and use of AI systems, but the lack of a comprehensive methodological framework limits its broader adoption. This paper addresses the gap by introducing the HCAI methodological framework (HCAI-MF), which includes five key components: HCAI requirement hierarchy, approach and method taxonomy, process, interdisciplinary collaboration approach, and multi-level design paradigms.

HCAI-MF, though conceptual, is a systematic and actionable methodology. It addresses current HCAI weaknesses and provides organizations with a methodological framework for HCAI practice effectively throughout the AI system lifecycle. To support adoption, this paper offers actionable recommendations to tackle potential implementation challenges, outlines future directions for evolving HCAI-MF, and introduces a "three-layer" strategy for integrating HCAI-MF across project, organizational, and societal levels.

As an emerging field, HCAI practice is at a critical juncture, with the potential to grow significantly, akin to the evolution of HCD over the past 40 years. This paper marks the beginning of a journey to refine and advance HCAI methodologies for advancing HCAI practice and shaping HCAI systems.

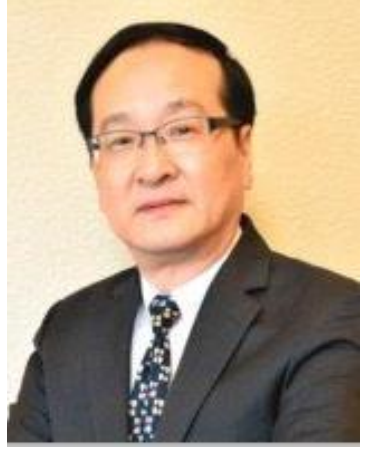

**Wei Xu** (SMIEEE) is a Professor of Zhejiang University, China. He is a Fellow of International Ergonomics Association, Human Factors and Ergonomics Society, and Association for Psychological Science. He received his Ph.D. in Human Factors and M.S. in Computer Science from Miami University in 1997. He serves as a Senior Editor for *IEEE TTS* and Associate Editor for *IEEE THMS*. His research areas include HCAI, HCI, and aviation human factors.

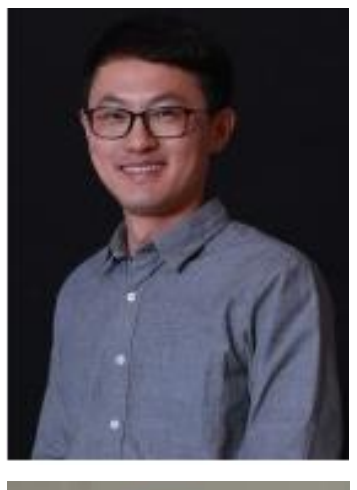

**Zaifeng Gao** is a Professor of Zhejiang University, China. He received his Ph.D. in Psychology from Zhejiang University, China, in 2009. His research interests include engineering psychology, autonomous driving, and cognitive psychology.

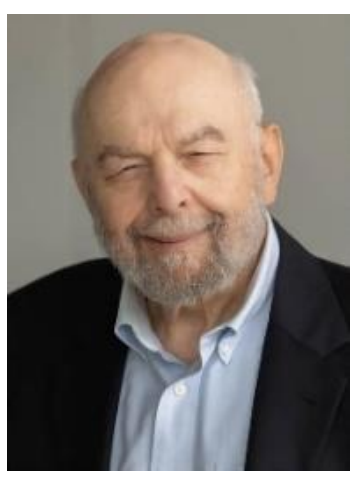

**Marvin J. Dainoff** received his Ph.D. in Psychology from University of Rochester in 1969. He is a Professor Emeritus of Miami University. He is a Past President of the Human Factors and Ergonomics Society. His research interests include sociotechnical system solutions, HCI, and ergonomics.

**Appendix A** Major HCAI approaches/methods and the human factors prioritized for human-centered AI systems

| Group | HCAI approaches & methods | Description | Human factors focused | Key HCAI-driven activities across the AI-based system development lifecycle stages | HCAI Guiding Principles Primarily Supported (refer to Table 2) |
|---|---|---|---|---|---|
| Human-centered strategy | Human value alignment with AI [68], [69], [70] | Creates AI systems that act in ways consistent with human values, ethics, and social norms, ensuring that AI systems, such as large language models (LLMs), make decisions and take actions that reflect human values, thereby reducing the risks of unintended harm. | Human values | Define human values. Ensure fairness in data collection. Mitigate bias in model development. Test with user feedback to align with human values and ethical performance. Deploy with clear communication to users. Monitor continuously with users and update as needed. | • Ethical Alignment<br>• Safety & Robustness |
| | Hybrid human-artificial intelligence [57], [71], [72] | Integrates the unique strengths of both human and artificial intelligence to enhance decision-making, problem-solving, and overall performance. In hybrid AI, humans contribute qualities like creativity, intuition, and ethical judgment, while AI provides data processing, pattern recognition, and computational speed. | Human strength | Define roles and complementary strengths. Collect data on task-specific inputs with human insights. Choose models optimized for collaboration and integrated human oversight with AI. Train/fine-tune using HITL for hybrid interactions. Test human-AI collaborative performance. Deliver & monitor collaboration for evolving user needs. | • Human Control & Empowerment<br>• Human Augmentation & Human-Led Collaborative Interaction |
| | Data and knowledge dual-driven AI [73], [74] | Integrates data-driven machine learning methods with knowledge-driven techniques (e.g., rules, expert systems). Integrates human expert knowledge into AI through technologies such as knowledge graphs, leveraging data and knowledge dual-driven AI to overcome LLM issues (e.g., hallucination, big data dependence, explainability). | Human knowledge | Define metrics and scenarios where human knowledge complements data insights. Acquire human knowledge (rules/expert insights). Select cognitive agent architectures to manage knowledge. Train with human feedback. Audit for ethical metrics. Use tools for users to verify the knowledge base. Gather user feedback to refine the knowledge base. | • Transparency and Explainability<br>• Ethical Alignment<br>• Human Augmentation & Human-Led Collaborative Interaction |
| Human-centered compu-ting | Human cognitive modeling [75], [76], [77] | Creates computational models that simulate human cognitive processes to enhance the design and functionality of AI systems. This approach ensures that AI systems are aligned with human thought processes, reasoning, and decision-making, fostering more intuitive and effective interactions between humans and AI. | Human cognitive capabilities | Identify goals and relevant cognitive tasks. Collect data on human behaviors & cognitive processes. Choose models simulating cognitive processes. Train models using real cognitive scenarios with user feedback. Assess alignment with users. Deploy UI for seamless cognitive interaction. Refine models to adapt to user needs. | • Human Augmentation & Human-Led Collaborative Interaction with AI<br>• User Experience |
| | Human-centered machine learning [21], [78] | Designs machine learning (ML) systems that prioritize human needs, values, and experiences throughout the AI system lifecycle, creating ML systems that are understandable, usable, fair, and aligned with users' expectations. It emphasizes transparency, interpretability, and human feedback loops, fostering trust and enhancing user experience. | Human values, needs, and feedback | User-centered problem definition, diverse data collection, explainable model selection, and user feedback integration. Usability testing with users and regular ethics audits uphold fairness and trust, while continuous monitoring and updates adapt models to evolving user needs and contexts. It is a user-participatory/interactive ML approach. | • Transparency and Explainability<br>• Ethical Alignment |
| | Human-centered explainable AI [19], [49] | Designs AI systems, such as LLMs, that prioritize transparency & interpretability. It aims to make AI decisions understandable and meaningful to users, enabling them to trust, interact with, and make informed decisions based on AI output by enhancing usability and aligning explanations with users' cognitive and emotional contexts. | Human knowledge, trust | Define user needs and domain knowledge levels. Select interpretable models. Validate explanations with human-centered metrics. Deploy interactive and transparent user interfaces with visualization. Continuous monitoring and updates maintain relevance to ensure accountability, empowering users and upholding ethical standards. | • Transparency and Explainability<br>• Ethical Alignment<br>• User Experience |
| Human-AI inter-action techno-logy and design | Intelligent user interface [79] | Focuses specifically on the design and functionality of UIs that leverage AI technologies to anticipate user needs and intentions, understand context and emotion, and respond dynamically to interactions (e.g., voice input). It centers around creating adaptive, responsive, personalized, intuitive, and supportive UI for users. | Human experience | Identify UX goals (e.g., accessibility, usability). Leverage diverse interaction data. Select appropriate UI/ multimodal technology, interaction model, and UI design paradigms to match user tasks and needs. Usability testing with user guides, refinements in evaluation, and deployment. Continuous monitoring/updates address user needs. | • User Experience<br>• Transparency and Explainability |
| | User experience design [55], [56] | Creates AI-driven interfaces and interactions that are intuitive, ethical, and aligned with users' needs. This approach considers not only usability but also transparency, interpretability, and trust, ensuring that AI systems are understandable and usable to users. | Human needs, experience | Conduct user research. Define UX requirements. Design explainable and intuitive UIs. Test with users and iteration improves UX, while personalization and accessibility make it adaptable and inclusive. Continuous monitoring with user inputs for improvements. | • Human Augmentation & Human-Led Collaborative Interaction<br>• User Experience |

| Group | HCAI approaches & methods | Description | Human factors focused | Key HCAI-driven activities across the AI-based system development lifecycle stages | HCAI Guiding Principles Primarily Supported (refer to Table 2) |
|---|---|---|---|---|---|
| | Human-AI interaction (e.g., robot, LLM) [80], [81], [82] | Takes a broader view of how humans and AI systems interact. It considers all aspects of interaction, communication, collaboration, user control, and ethical implications using AI technology. Aims to create holistic interactions for humans to interact with AI, maintain oversight, trust, and human control. It includes humans interacting with AI systems (e.g., robots, XR/metaverse, wearables, LLM). | Human experience, trust, collaboration | Define goals and diverse interaction context. Develop multimodal interaction technology by leveraging AI-based approaches (e.g., intent recognition) and appropriate interaction design paradigms and models. Build explainability to foster transparency and trust. Personalization and adaptability are integrated to meet individual preferences, supported by rigorous testing for usability and performance. | • Human Augmentation & Human-Led Collaborative Interaction<br>• User Experience |
| | Human-led collaborative interaction with AI [11] | Designs AI systems to ensure humans guide and oversee AI in a partnership, guaranteeing that AI augments rather than replaces human judgment. Humans leverage AI's computational strengths while retaining decision-making control, aligning AI outputs with human values, context, and goals | Human leadership role | Set human/AI roles and collaborative goals. Define human-AI collaboration models (e.g., shared tasks, situation awareness, and responsibilities) with humans as leaders. Evaluation ensures trust, human controllability, and usability. Deployment provides seamless UI, with monitoring/updates refining trustworthy collaboration. | • Human Augmentation & Human-Led Collaborative Interaction<br>• User Experience |
| Human-contro-llability over AI | Human-in/on-the-loop system design [83], [84], [85], [86] | Designs AI systems to integrate humans into the AI decision-making process, allowing for real-time input, intervention, and oversight, ensuring that AI systems can benefit from human judgment, adapt to nuanced contexts, and improve based on human feedback. It is valuable in high-stakes scenarios requiring human expertise. | Human role, controllability | Define human intervention tasks and roles. Choose models that can adapt to human input in real-time. Test and refine models with human corrections, and validate models with human-in/on-the-loop workflows. Deployment enables user-intuitive controls and override options, while monitoring leverages user feedback to improve workflows. | • Human Control & Empowerment<br>• Ethical Alignment |
| | Machine behavior management [87] | Monitors and regulates AI autonomous behaviors to align with human goals, ethical and safety requirements. Designs machine monitoring mechanisms, sets constraints, and enables human intervention when needed. It is essential for preventing unintended outcomes and promoting responsible AI. | Human ethics | Define AI machine behaviors for ethical and reliable actions. Use diverse training data and model-controlled responses. Testing, real-time monitoring, continuous feedback, and regular audits maintain alignment with ethical standards, while safe decommissioning ensures ethical accountability. | • Ethical Alignment<br>• Human Control & Empowerment |
| | Meaningful human control [88], [89] | Designs AI with humans' oversight, influence, and accountability, particularly in key decision-making cases. It ensures that humans can understand, guide, and, if needed, intervene in AI systems to prevent unintended harm for ethical standards and societal values. It ensures AI remains transparent/predictable to align with human intentions. | Human accountability | Define roles and metrics for intervention. Data collection focuses on scenarios requiring judgment. Design models for explainability, real-time alerts, and accountability. Training/testing uses feedback to refine control mechanisms and system ability with human intentions. Deploy an intuitive UI for monitoring/intervention, with continuous updates. | • Human Control & Empowerment<br>• Ethical Alignment<br>• Accountability |
| Human-centered gover-nance | Ethical AI standards and governance [43], [62], [90] | Develops standards, processes, and practices for ethical development, deployment, and management of AI systems, ensuring they align with societal values, ethics, and legal standards. This focuses on accountability, transparency, fairness, and safety, addressing concerns like bias, privacy, and unintended harm. | Human values, accountability | Define and follow ethical AI standards with stakeholder engagement. Data collection focuses on bias mitigation, while model development ensures ethical and transparency. Testing includes ethics audits for compliance, and deployment emphasizes clear communication. Continuous monitoring and user feedback maintain ethical alignment. | • Ethical Alignment<br>• Safety & Robustness |
| | Algorithm governance [91], [92] | Establishes frameworks, policies, and practices to oversee the development, deployment, and use of algorithms, ensuring they operate transparently, ethically, and in alignment with human values. This sets guidelines for accountability, fairness, bias mitigation, and privacy in algorithmic decision-making processes for managing risk. | Human ethics | Define ethical and operational goals for algorithms. Collect quality data to ensure privacy and bias mitigation. Select algorithms with built-in fairness and bias mitigation. Training with user input to ensure algorithms are aligned with ethical goals. Test compliance and societal impact. Deployment includes user feedback with ongoing monitoring. | • Ethical Alignment<br>• Safety & Robustness |
| | Data governance [53], [54] | Manages data specifically used in AI systems, focusing on its quality, security, ethical use, and compliance with legal standards. It encompasses policies and practices to ensure the data is accurate, unbiased, and protected, facilitating responsible and transparent AI. It helps manage risks and promotes trust in AI decision-making. | Human ethics, privacy | Define goals (e.g., quality, privacy). Collect data to ensure ethical issues (e.g., biases, fairness). Model design uses privacy-preserving techniques and usage logs. Training uses secure, quality data aligned with policies. Audit adherence to standards. Deploy safeguards and access controls. Continuous monitoring addresses quality and security. | • Ethical Alignment<br>• Safety & Robustness |

Appendix B Key HCAI activities and methods across the AI life cycle for delivering human-centered LLM

| **LLM lifecycle** | **Typical Activities across LLM lifecycle** | **HCAI Challenges in the l LLM Lifecycle (examples)** | **Recommended HCAI Methods/Activities (examples) across LLM Lifecycle based on the HCAI Process** |
|---|---|---|---|
| Problems & Requirements | Define functional requirements for model output, accuracy, and speed. | *Lack of User Involvement*: Limited consideration of diverse users and societal impacts leads to models misaligned w/ human values/needs. | Involve users and stakeholders in defining human-centered objectives (e.g., ethics, safety, social impact) in the early stages; set values-based goals (e.g., transparency, fairness) and UX goals (e.g., UI, prompt adaptability, accessibility) [102], [103], [104]. |
| Data Collection & Preparation | Collect and preprocess large datasets, ensuring they are representative of general language. | *Bias & Privacy Issues*: Collection methods may embed societal biases, leading to harmful outputs. Data privacy is often overlooked. | Prioritize inclusive, fair data from diverse perspectives; remove harmful content with human oversight; address privacy issues upfront; ensure balanced data across multimodal modes (e.g., text, image) to avoid biases and inconsistencies [101], [105]. |
| Model Selection & Architecture Design | Choose scalable architecture and focus on performance and memory efficiency. | *Interpretability*: Architectures often lack transparency, reducing trust and making it difficult for users to understand how decisions are made. | Select architectures for explainability and integrate human-in-the-loop (HITL) or human-on-the-loop (HOTL) oversight; incorporate capabilities for prompt customization and multimodal input for UX and adaptability for diverse users [106], [107]. |
| Training & Fine-tuning | Pre-train & fine-tune on data for high accuracy. Implement RLHF to align with basic ethical norms. | *Ethical Risk Mitigation:* Standard fine-tune may not address all ethical issues, especially around unintended biases. Limited alignment with human values. | Use human feedback and ethical evaluations training/fine-tuning. Focus on continuous validation/audits for unintended outputs. Align with human values, consistent performance, and alignment across multimodality), such as instruction tuning and supervised fine-tuning [108], [109], alignment tuning, reinforcement learning from human feedback [110], [111]. |
| Evaluation & Validation | Evaluate standard technical benchmarks. Use limited adversarial testing for vulnerabilities. | *Limited Real-World Testing*: Standard metrics don't account for interpretability, UX or ethical impact. Limited adversarial tests can miss real-world ethical risks. | Conduct iterative testing with real users with user-centered metrics (e.g., interpretability, relevance, and ethical impact), including prompt response quality, UI usability, and consistency and fairness across multimodal. Techniques include model cards [112], human-centered auditing [113], algorithmic auditing [114]. |
| Deployment | Deploy with emphasis on technical performance (e.g., speed, reliability) and API usability. | *Limited Transparency & Control:* Deployments lack transparency, users often cannot control or understand outputs, limiting trust. | Include user-friendly interfaces, explainability, and continuous monitoring; provide options for prompt customization, user feedback, and transparency in multimodal outputs, allowing users to interact across text, image, and other supported modes [34]. |
| Monitoring & Improvement | Track performance to identify drifts. Update the model for quality. Focus on technical costs. | *Bias Persistence & User Feedback:* Limited user feedback make it challenging to identify and correct biases or ethical issues over time. | Engage users for feedback, monitor ethical issues, and address cultural sensitivity; include transparent communication with users about updates, any changes in prompt responses or multimodal performance; validate updates with users [115], [116]. |

Appendix C Needs and Gaps Analysis of human-AV co-driving using the HCAI multi-level design paradigms

| **HCAI design paradigms** | **Aspects of analysis** | **Gaps and challenges in Current approaches** | **High-level needs to address the challenges based on HCAI approaches and methods** |
|---|---|---|---|
| Human-AI joint cognitive systems | Collaborative relationship, dynamic role allocation | SAE [134-131] defines six numerical classifications of automated driving, a technology-centered taxonomy that focuses on the binary function allocation between drivers and AVs. Consider AVs as tools or replace drivers [145]. | Advocate for a collaborative relationship and dynamic functional allocation between drivers and AVs (AI agents). Consider AVs as *cognitive agents* collaborating with drivers, forming a human-AI joint cognitive system. Adopt a "human-in/on-the-loop" design. |
| | Human-centered approach | Take a more technology-centered design approach, leaving an unclear leadership role of human driver-AV systems [146], [147]. | Adopt the "human-centered" approach, promoting human-led collaboration (L4 & below) and the final controllers. Promote meaningful human control for humans to oversight and accountability. |
| | Driver-AV complementarity design | Ineffective leveraging of the strengths of humans and AI with unbalanced automation and human control, lack of dynamic role adaptation, and lack of collaborative & adaptive UI to facilitate human-AI complementarity [148]. | Structure the AV system to augment human drivers and retain AV's role as a collaborator, instead of replacing them entirely. Balance automation and human control with adaptive UI to leverage the strengths of both humans & AI to maintain human oversight/control. |
| | Collaborative interaction design paradigm | Adopt a unidirectional interaction paradigm that is primarily based on the traditional "stimulus-response" unidirectional human-machine interaction, resulting in non-collaborative UI [149]. | Adopt a collaborative design paradigm with bi-directional interaction between drivers and AVs. AI monitors driver physiological, cognitive, and intentional states; drivers obtain optimal situation awareness to remain engaged and facilitate effective human takeover if needed. |
| Human-AI joint cognitive ecosystems | Ecosystem perspective | Focus on individual human-AV systems. Such a narrow focus cannot guarantee the optimized design and safe operations of the entire AV ecosystem [40], [142], [143]. | Extend to a human-AV ecosystem perspective: a human-AV-pedestrian-road-traffic infrastructure system, forming a joint cognitive ecosystem for the collaboration of all agents within it. |
| | Collaboration across human and AI agents | Do not consider the interaction and collaboration across agents within the co-driving ecosystem, e.g., drivers, pedestrians, AI agents within AVs, AI-based road systems, and AI-based traffic command systems [150]. | Consider that all humans and AI agents are cognitive agents within the ecosystem. System design addresses distributed collaboration, trust, interaction, learning, situation awareness, and control across all agents within the ecosystem (e.g., AV-AV interaction, road right allocation). |
| | Human-centered design | Achieve the "human-centered AI" design philosophy at the level of individual drive-AV systems, and cannot guarantee a human-centered co-driving ecosystem [146]. | Achieve HCAI design at an ecosystem level to ensure humans retain ultimate control, e.g., AVs losing control can be taken over by drivers, operation centers, or traffic command centers (e.g., 5G remote control). |
| | Ecological learning & evolving | Do not fully consider the cognitive capabilities of AI agents, resulting in insufficient consideration of the learning, evolving, and adaptive features of all agents within the co-driving ecosystem [151]. | Design for co-learning, co-evolution, and co-adaptive capabilities of all agents. Establish a trusted ecosystem across agents and develop effective compatibility design mechanisms to address conflicting ethical rules across AV systems, achieving effective collaboration. |

| Intelligent sociotechnical systems | Sociotech-nical systems perspective | Focus on the technical aspect of the co-driving system. Do not fully consider the impacts of non-technical aspects from a sociotechnical systems perspective, such as human ethics, public awareness and acceptance, and driver behavioral changes [40], [143], [144]. | Consider co-driving systems to be a sociotechnical system. Consider the joint optimization between AV technical and social subsystems to ensure AVs are socially accepted and aligned with ethical/social rules. The overall safety of the co-driving system relies on the joint optimization between technical and non-technical subsystems. |
|---|---|---|---|
| | Human-centered approach | Do not fully consider building trust and managing shared control between humans and AVs to establish ethical & safety protocols for decision-making, compliance with regulations. Lack of integration for AVs to meet technical demands, human values, & social expectations [152]. | Create AV systems that prioritize human needs and safety while integrating social and technical elements. Recognize AVs do not operate in isolation and exist within sociotechnical environments involving interactions between humans, technology, infrastructure, and regulations. Ensure humans are the ultimate decision makers |
| | Social integration & behavioral adaptation | Low public awareness and trust issues, lack of clear communication of safety measures, and technology benefits. Do not fully consider behavioral changes and adaptations for all road users. Policies do not ensure AV benefits reach a broad demographic [153]. | Understand and address broader social factors (e.g., public trust, cultural norms, social acceptance) to ensure AV systems are aligned with human values, societal acceptance, and trust in AV technologies. Foster AV accessibility in technology for equitable deployment. Address driver upskilling and certification in AVs. |
| | Harmonized ethical & safety standards | Ethical dilemma issues (e.g., prioritizing lives in an accident). Lack of value-laden decision-making algorithms. Accountability issues for the responsibility of accidents. No standards ensuring AVs can safely handle diverse scenarios [154]. | Harmonize technologies with safety boundaries, protocols, and ethical standards. Guide AV decision-making of potential harm (e.g., passenger vs. pedestrian). Clarify liability in accidents and responsibilities (e.g., driver vs. vendor). Ensure AV systems can respond to various human behaviors and environmental factors. |
| | User data privacy, security, & governance | Lack of data privacy as AVs gather user data, Unclear data ownership and usage issues. Cybersecurity issues due to vulnerability to cyber-attacks endangering passengers and road users [155]. | Develop data privacy protocols to protect data collected by AVs. Define data ownership rights for users, manufacturers, and third parties. Establish cybersecurity protocols to prevent potential hacking through system updates & multi-layered security measures. |
| | Multi-stakeholder collaboration | Lack of effective collaboration among automakers, government, and urban planners. Lack of global standards to facilitate AV adoption worldwide. Lack of consumer protection and liability [156], [157]. | Create consistent regulations to enable smooth cross-border operations. Establish rigorous certification standards to ensure AVs meet safety, ethical, and environmental benchmarks. Modify insurance and liability policies to address emerging AV risks. |

## Appendix D Mapping of HCAI-MF components for human-AV co-driving solutions

| **HCAI process and HCAI interdisciplinary collaboration approach** → | | | |
|---|---|---|---|
| **HL Design themes based on the design paradigm** (Figure 6) | **Design actions based on HCAI principles** (Table 2) | **Approaches & methods to be applied** (Appx A) | **HCAI-based high-level requirements (examples) driven by the HCAI design paradigm, guiding principles, and approaches & methods** |
| *Human-AI joint cognitive systems:*<br>Collaborative driver-AV relationship, dynamic role allocation<br>Human-centered approach<br>Driver-AV complementarity design<br>Collaborative interaction design | *Human control & empowerment:* Allow humans to understand, influence & control AI when needed<br>*User experience:* Create interactions that are engaging, intuitive, accessible, and aligned with user expectations.<br>*Transparency & explainability:* Provide explainable & understandable AI output for humans<br>*Human-led collaboration & augmentation:* Design AI to enhance human abilities and human-led collaboration | *Human-in/on-the-loop design:* (drivers) oversight and take control<br>*Intelligent user interface:* Design collaborative UI<br>*User experience design:* Build a good experience<br>*Human state recognition modeling:* Model and sense driver states<br>*Human-AI interaction:* Use multimodal UI<br>*Human-led collaboration AI*: Design driver-led collaboration<br>*Human-centered explainable AI:* Make AV decision explainable | *Dynamic & collaborative role allocation*: Define collaborative roles for drivers and AVs under various scenarios for smooth transitions of AV control. Drivers supervise high-level operations (e.g., emergency), and systems provide comprehensive situation awareness for effective monitoring and intervening when anomalies are detected. AV handles dynamic driving tasks. |
| | | | *Shared situation awareness*: Maintain shared understanding of the environment. Keep the driver in or on the loop, drive oversight with good situation awareness, and take over control if needed. Provide situation awareness displays to visualize the AV's decisions, context, and potential risks. Offer context-relevant output to assist human supervisors in making informed interventions. |
| | | | *Transparency & explainability:* Explain why the AV is performing specific actions through visual, auditory, or haptic feedback to maintain driver trust (e.g., "changing lanes to avoid a slow-moving vehicle"). Use displays or auditory prompts to inform the driver about the AV's confidence levels and limitations for transparent decision-making. |
| | | | *Collaborative decision-making:* Facilitate collaborative decision-making for drivers to approve, reject, or modify AV decisions. Design for easy overriding of AV actions when required (e.g., AV suggests an alternate route for drivers to accept or reject). Provide two-way communication to ensure the AV communicates its current state and decisions (e.g., changing lanes in 5 sec). |
| | | | *Collaborative monitoring and adaptation:* Deploy driver state monitoring by using sensors to monitor driver attention and readiness (e.g., eye tracking, grip sensors). Adapt AV behavior based on the driver's state (e.g., take full control if the driver is distracted). Use continuous learning to allow the AV to learn/adapt to individual driver preferences and behaviors over time. |
| | | | *Collaborative and adaptive interaction:* Use adaptive UI to support varying levels of human involvement. Adjust AV's level of autonomy based on driver behavior (e.g., systems prompt driver engagement if reduced attention is detected, taking over when drivers are unresponsive). |
| | | | *Collaborative user interfaces:* Design intuitive UI to facilitate collaboration. Prioritize critical information and avoid cognitive overload by presenting the most relevant information. Use multimodal UI to incorporate visual (e.g., head-up displays), auditory (e.g., voice alerts), and haptic (e.g., vibrations) real-time feedback for effective communication to drivers. |
| | | | *Collaborative emergency responses*: Enable collaborative handling of critical situations (e.g., AV performs an emergency stop if the driver does not respond). Mitigate risk by predicting potential emergencies and alerting the driver early. Provide a smooth handoff for the driver to take control. Ensure AV maintains temporary control if the driver fails to respond. |